\documentclass[aps,pre,twocolumn,superscriptaddress,showpacs,floatfix]{revtex4-1}
\usepackage{dcolumn}
\usepackage{amsmath}
\usepackage{amssymb}
\usepackage{amsthm}
\usepackage{graphicx}
\usepackage{bm}
\usepackage[T1]{fontenc}
\usepackage{color}
\usepackage{url}
\usepackage[bf]{subfigure}
\usepackage{rotating}
\usepackage{scalefnt}
\usepackage{multirow}

\usepackage{scalefnt}
\usepackage{times}
\usepackage{mathtools}
\usepackage{bbm}

\usepackage{tikz}
\usetikzlibrary{topaths,calc}

\theoremstyle{definition}

\newcommand{\J}{\mathbf{J}}

\DeclareMathOperator{\I}{\mathbf{I}}
\DeclarePairedDelimiter{\ceil}{\lceil}{\rceil}

\newcommand{\vertii}[1]{{\left\vert\kern-0.25ex\left\vert #1 \right\vert\kern-0.25ex\right\vert}}

\newcommand{\E}[1]{\mathbb{E} \left( #1 \right)}
\newcommand{\Prob}[1]{\mathbb{P}\left( #1 \right)}

\begin{document}

\title{Social contagion models on hypergraphs}

\author{Guilherme Ferraz de Arruda}
\affiliation{ISI Foundation, Via Chisola 5, 10126 Torino, Italy}

\author{Giovanni Petri}
\affiliation{ISI Foundation, Via Chisola 5, 10126 Torino, Italy}

\author{Yamir Moreno}
\affiliation{Institute for Biocomputation and Physics of Complex Systems (BIFI), University of Zaragoza, 50018 Zaragoza, Spain}
\affiliation{Department of Theoretical Physics, University of Zaragoza, 50018 Zaragoza, Spain}
\affiliation{ISI Foundation, Via Chisola 5, 10126 Torino, Italy}

\begin{abstract}
Our understanding of the dynamics of complex networked systems has increased significantly in the last two decades. However, most of our knowledge is built upon assuming pairwise relations among the system's components. This is often an oversimplification, for instance, in social interactions that occur frequently within groups. To overcome this limitation, here we study the dynamics of social contagion on hypergraphs. We develop an analytical framework and provide numerical results for arbitrary hypergraphs, which we also support with Monte Carlo simulations. Our analyses show that the model has a vast parameter space, with first and second-order transitions, bi-stability, and hysteresis. Phenomenologically, we also extend the concept of latent heat to social contexts, which might help understanding oscillatory social behaviors. Our work unfolds the research line of higher-order models and the analytical treatment of hypergraphs, posing new questions and paving the way for modeling dynamical processes on these networks.
\end{abstract}

\maketitle

Network science has had a radical impact on our knowledge about critical dynamics on complex systems. This is particularly true when it comes to inspect social and biological contagion processes~\cite{Barrat08:book,Newman010:book,Boccaletti06:PR,Vespignani2015,Arruda2018}, where new and relevant phenomenologies have been discovered~\cite{Vespignani2000,Goltsev2012,Vespignani2015,Arruda2018}. For instance, while classical spreading models predict finite critical points~\cite{Barrat08:book,Vespignani2015,Arruda2018}, heterogeneous networks often present vanishing transitions~\cite{Barrat08:book,Vespignani2000,chatterjee2009,Vespignani2015,Arruda2018}, supporting the predictions in real world networks~\cite{Colizza2007,Balcan2009,Tizzoni2012,Zhang2017}. Theories of contagion covered many aspects, from different contagion types to richer substrates underlying the process itself. 
A particularly relevant development is the extension of contagion processes to multilayer networks, which in turn paved the way to combinatorial higher-order models. Indeed, multilayer's structural~\cite{Kivela2014,Arruda2017,Cozzo2018,Arruda2018njp,Aleta2019} and spreading and diffusion properties~\cite{Gomez2013,Kivela2014,Arruda2017,Arruda2018} have a new and richer phenomenology. Nevertheless,  as recently argued in~\cite{Lambiotte2019}, real data is revealing that pairwise relationships -- the fundamental interaction units of networks -- do not capture complex dependencies. For instance, modern messaging systems (e.g., WhatsApp, Telegram, Facebook Messenger, among others) allow users to communicate in groups, which create a direct channel for information diffusion among all members of that given group. In other words, modern information spreading is often a one-to-many process. In the same way, team collaborations are inherently group interactions, as are some types of molecular interactions  \cite{benson2018simplicial}. Moreover, the sizes of such groups can be very different, spanning orders of magnitude. Thus, a graph-based approach might not be sufficient to describe systems that involve interactions over many different scales and orders. Evidence from social and biological studies provided initial indications that such interactions can have crucial effects \cite{Iacopo2019,Caetano2019,Centola2018}. Understanding their properties and effects is therefore of paramount importance.

Combinatorial higher-order models~\cite{Lambiotte2019} offer a way to describe these systems, by overcoming some of the limitations of classical, lower-order network models. In a first attempt, Iacopini et al.~\cite{Iacopo2019} presented a model of social contagion defined on simplicial complexes and provided approximate solutions for complexes of order three, including new phenomenological patterns associated to the critical properties of the dynamics. However, the proposed model is still very constrained, both structurally and dynamically. Here we adopt hypergraphs, which relax the structural restrictions required by simplicial complexes by imposing virtually no limitation on the type, size and mutual inclusion of interactions, thus, representing more faithfully and naturally real systems. We further incorporate explicit critical-mass dynamics, which generalizes the one modeled in~\cite{Iacopo2019}. We report analytical and numerical analyses of the theoretical framework introduced here as well as results for several limiting cases and hypergraph structures. In doing so, we uncover the presence of discontinuous transitions and bistability led by higher-order interactions and critical-mass dynamics. These transitions contrast with classical contagion models on complex network, which instead display continuous transitions, e.g., SIS or SIR disease spreading. The resulting model thus displays a rich complex phenomenology, remaining very flexible, and able to cover a wide range of systems. We round off the paper by discussing several implications of our study, and most notably, the role of critical mass dynamics in social contagion, providing new insights that could help explaining reported differences in experimental results ~\cite{Kanter1977, Drude1988, Grey2006, Centola2018}.

\begin{figure}[t]
\includegraphics[width=\linewidth]{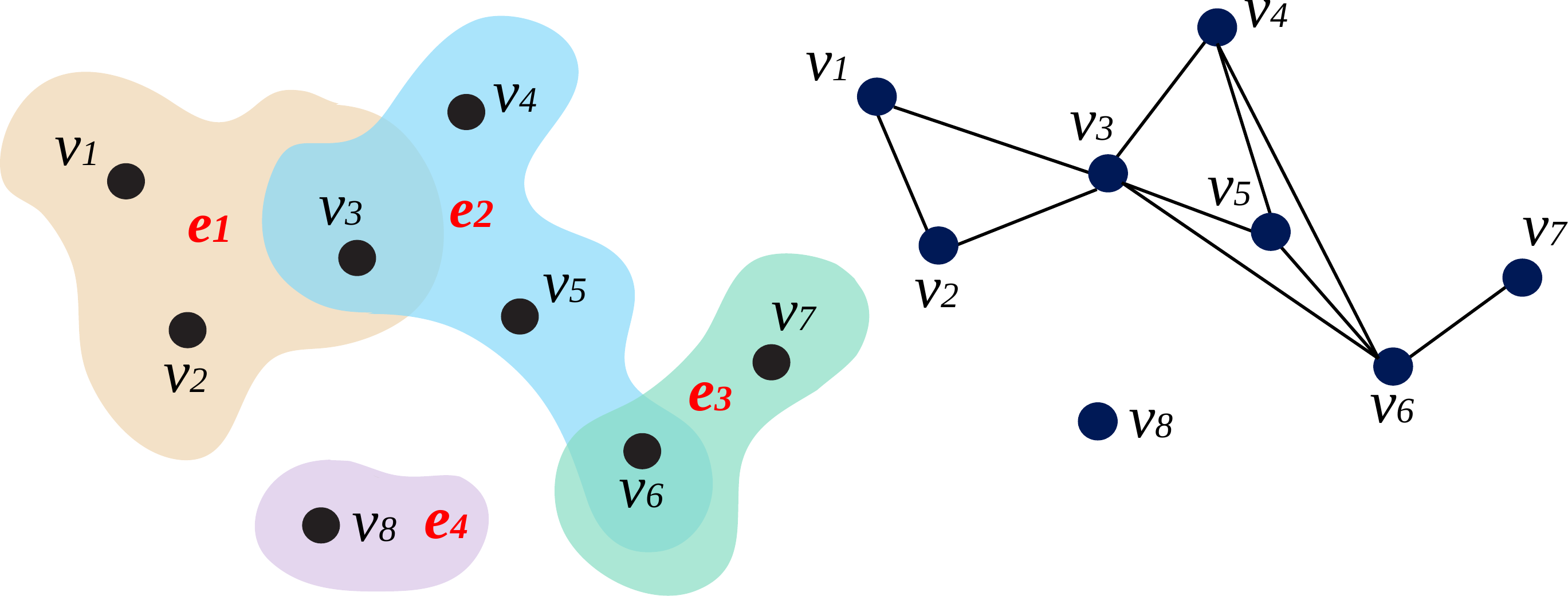}
\caption{Graphical representation of a hypergraph. Mathematically, $\mathcal{V} = \{v_1, v_2, v_3, v_4, v_5, v_6, v_7, v_8 \}$, $\mathcal{E} = \{e_1, e_2, e_3, e_4 \}$, where the hyperedges are $e_1 = \{v_1, v_2, v_3 \}$, $e_2 = \{v_3, v_4, v_5, v_6 \}$, $e_3 = \{v_6, v_7 \}$ and $e_4 = \{v_8 \}$. On left we have the hypergraph and, on right, the graph projection, where hyperedges are simplified as $|e_j|$-cliques.}
\label{fig:hypergraph_ex}
\end{figure}

Let us first introduce some formal definitions. A hypergraph is defined as a set of nodes, $\mathcal{V} = \{ v_i \}$, where $N = |\mathcal{V}|$ is the number of nodes and a set of hyperedges $\mathcal{E} = \{ e_j \}$, where $e_j$ is a subset of $\mathcal{V}$ with arbitrary cardinality $|e_j|$. If $\max \left( |e_j| \right) = 2$ we recover a graph. On the other hand, if for each hyperedge with $|e_j| > 2$ its subsets are also contained in $\mathcal{E}$, we recover a simplicial complex (for more on the hypergraph structure, see Supplementary Material, Section~\ref{sec:structure}).
Fig.~\ref{fig:hypergraph_ex} shows an example of a hypergraph and its graph projection. In an arbitrary hypergraph, we associate to each individual $v_i$ a Bernoulli random variable $Y_i$ (complementary $X_i$). If the node $v_i$ is active $Y_i = 1$ ($X_i = 0$), otherwise $Y_i = 0$ ($X_i = 1$). To each active node, we associate a deactivation mechanism, modeled as a Poisson process with parameter $\delta_i$, $N_i^{\delta}$ ($Y_i \xrightarrow[]{\delta_i} X_i$). 
For each hyperedge, $j$, we define a random variable $T_j = \sum_{k \in e_j} X_k$, which is the number of active nodes in the hyperedge. 
If $T_j$ is equal or above a given threshold, $\Theta_j$, we associate a Poisson process with parameter $\lambda_j$, $N_j^{\lambda_j}$ (that is, if $T_j \geq \Theta_j$, then $X_k \xrightarrow[]{\lambda_j} Y_k$, $\forall k \in e_j$). In other words, the dynamics is given by a threshold process that becomes active only above a critical mass of activated nodes. 
Moreover, if $|e_j| = 2$, we assume that the Poisson processes are directed, implying that it is not a threshold process anymore. 
This definition allows recovering traditional SIS contagion models. While the proposed model is general in that it allows for arbitrary heterogeneity in parameters, we focus on more straightforward, but representative, cases. We assume that $\delta_i = \delta$ and $\lambda_j = f(|e_j|)$, where $f$ is an arbitrary function of the cardinality of the hyperedge. It is also convenient to define $\Theta = \ceil[\big]{\Theta^* N}$, where $\Theta^*$ is a real number representing the fraction of active nodes.

The exact equation that describes the aforementioned dynamics can be written as
\begin{equation} \label{eq:exact}
\begin{split}
 &\dfrac{d \E{Y_i}}{dt}= \\
 &\E{-\delta Y_i + \left( 1 - Y_i\right) \sum_{e_j\cap \{v_i\} \neq \emptyset} \sum_{k \in \{e_j \backslash v_i\}} \lambda_j \mathbbm{1}_{\{ (T_j - Y_k) \geq \Theta_j\}}},
\end{split}
\end{equation}
where the first summation is over all hyperedges containing $v_i$, and the second over all the neighbors in that hyperedge. 
Furthermore, $\mathbbm{1}_{\{ T_j - Y_i \geq \Theta_j\}}$ is an indicator function that is $1$ if the critical mass in the hyperedge is reached, and $0$ otherwise.  Naturally, the order parameter is defined as the expected fraction of active nodes, i.e., $\rho = \frac{1}{N} \sum_i \E{Y_i}$. 

Although Eq.\ref{eq:exact} captures the exact process, it cannot be numerically solved. Thus, assuming that the random variables are independent and denoting $y_i = \E{Y_i}$, we obtain the first-order approximation, given as
\begin{equation} \label{eq:first_order}
  \dfrac{d y_i}{dt}= -\delta y_i + \lambda \left( 1 - y_i\right) \sum_{e_j\cap \{i\} \neq \emptyset} \sum_{k = \Theta_j}^{|e_j|} \lambda^* (|e_j|) \mathbb{P}_{e_j} \left(K=k \right),
\end{equation}
where we assume that the spreading rate is composed by the product of a free parameter and a function of the cardinality, i.e., $\lambda_j = \lambda \times \lambda^*(|e_j|)$. In this formulation, we estimated the expectation of the indicator function as a Poisson binomial distribution (for more on this approximation, see SM, Section~\ref{sec:F}). Formally,
\begin{eqnarray}
 \E{\mathbbm{1}_{\{ (T_j - Y_k) \geq \Theta_j\}}} \approx \sum_{m = \Theta_j}^{|e_j|} \mathbb{P}_{e_j} \left(K=l \right) \\
 \mathbb{P}_{e_j} \left(K=l \right) = \sum\limits_{A\in F_l} \prod\limits_{i\in A} y_i \prod\limits_{j\in A^c} (1-y_j), \label{eq:prob_pb}
\end{eqnarray}
where $F_l$ is the set of all subsets of $k$ integers from $\{1, 2, ... n=|e_j|\}$,  $A$ is one of those sets, and $A^c$ is its complementary. 
Intuitively, $A$ accounts for the possibly active nodes and $A^c$ the possibly inactive ones. Thus, the summation over $F_l$ considers all possible configurations in a given hyperedge. Equation~\ref{eq:prob_pb} is not numerically stable if $|e_j|$ is large. It is however possible to stabilize its solutions by considering the discrete Fourier transform~\cite{Fernandez2010}
\begin{equation} \label{eq:dft_pn}
 \mathbb{P}_{e_j} \left(K=k \right) = \frac{1}{n+1} \sum\limits_{l=0}^n C^{-lk} \prod\limits_{m=1}^n \left( 1+(C^l-1) y_m \right),
\end{equation}
where $C=\exp \left( \frac{2i\pi }{n+1} \right)$, which then allows to compute the solution for arbitrarily large hyperedges. Interestingly, although the whole argument is quite intricate, Eq.~\ref{eq:dft_pn} is simple, allowing the numerical evaluation of Eq.~\ref{eq:first_order}.

\begin{figure*}[t]
\includegraphics[width=\linewidth]{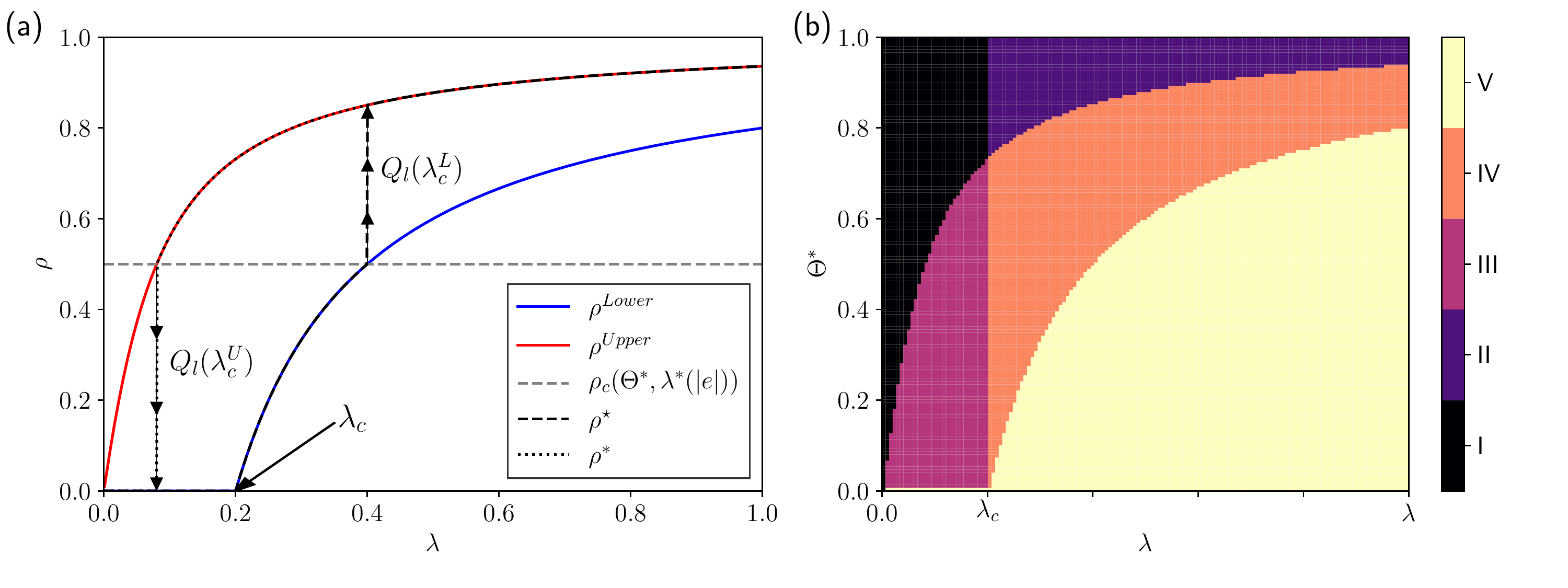}
\caption{Results for the hyperblob. Panel (a) shows the possible solutions for a fixed $\Theta^*=0.5$. In red and blue, the upper and lower solutions (branches), respectively. The transition from the lower to the upper solution (upper to lower) occurs at the intersection of the lower (upper) solution with a value of $\rho_c$ in which the upper solution became stable (unstable). The discontinuity is characterized by the latent heat, $Q_l(\lambda_c^{L})$ or $Q_l(\lambda_c^{U})$. At $\lambda_c = 0.2$, the lower solution shows a second-order phase transition. In (b) Schematic of the parameter space: Region I: the absorbing state for both the lower and upper solution; Region II:  only the lower solution is stable (the global critical mass is not reached, $\rho < \rho_c$); Region III: $\rho^{Upper}$ is stable and $\rho^{\text{Lower}} = 0$ (below the critical point); Region IV: $\rho^{Upper} > \rho^{Lower} > 0$ and both are stable (bi-stable); Region V: only the upper solution is stable (the global critical mass was reached, $\rho \geq \rho_c$).}
\label{fig:schematic}
\end{figure*}

Our main result is that contagion on hypergraphs is characterized by a rich and diverse phase-space, generally populated by continuous and discontinuous transitions and hysteretic behaviors. In particular, we have analytically observed discontinuity and bi-stability in the order parameter on top of some regular structures. We provide full details of the calculations in the SM (see sections~\ref{sec:hyperblob} and~\ref{sec:Hyperstar}) for two limiting cases, namely, a hypergraph composed by a hyperedge containing all nodes in addition to a random regular network (which we call hyper-blob), and a star (referred to as hyper-star). For the sake of clarity, let us show the main results for the hyperblob. For this case, we can exploit the structural symmetries to solve $\rho(\lambda, \lambda^*, \delta)$, obtaining two locally stable solutions. Specifically, consider a hypergraph built up as a homogeneous set of pairwise interactions with average degree $\langle k \rangle$ and a single additional hyperedge containing all nodes. In this case, the order parameter can be solved as 
\begin{eqnarray}
& &\rho^{\text{Lower}} = 
 \begin{cases}
  1 - \frac{\delta}{\langle k \rangle \lambda}, \hspace*{1cm} \text{if} \hspace*{2mm} \frac{\lambda}{\delta} \geq \frac{1}{\langle k \rangle}\\
  0, \hspace*{2.1cm} \text{otherwise}
 \end{cases}\\
  & &\resizebox{.9\hsize}{!}{$\rho^{\text{Upper}} = \frac{-\delta + \langle k \rangle \lambda - \lambda^* \lambda + \sqrt{ 4 \langle k \rangle \lambda^* \lambda^2 + (\delta + (-\langle k \rangle + \lambda^*) \lambda)^2}}{(2 \langle k \rangle \lambda)},$}
\end{eqnarray}
where a second-order phase transition for $\rho^{\text{Lower}}$ is naturally obtained as $\frac{\lambda}{\delta} \geq \frac{1}{\langle k \rangle}$ ~\cite{Arruda2018}. Furthermore, the discontinuities can also be calculated as 
\begin{eqnarray}
 \lambda_c^{\text{L}} &=& \frac{\delta}{\langle k \rangle - \Theta^*  \langle k \rangle} \\
 \lambda_c^{\text{U}} &=& -\frac{\delta  \Theta^* }{\lambda^*  \Theta^* -\lambda^* +(\Theta^*)^2 \langle k \rangle - \Theta^* \langle k \rangle}.
\end{eqnarray}
Phenomenologically, a discontinuity implies that our system possesses a ``social latent heat'', that is released or accumulated at a constant value of $\lambda$. More specifically, before the discontinuity, ``energy'' has been stored in the partial activation of the hyperedges. At the discontinuity this ``energy'' is absorbed (released) at once for a constant value of $\lambda$. In fact, the social latent heat can be expressed as
\begin{equation}
 Q_l(\lambda_c^{X}) = \left( \rho^{\text{Upper}}(\lambda, \delta, \lambda^*, N) - \rho^{\text{Lower}}(\lambda, \delta, \lambda^*, N) \right)_{\lambda = \lambda_c^X},
\end{equation}
where $Q_l(\lambda_c^{X})$ can be $Q_l(\lambda_c^{\text{L}})$ (energy absorbed) or $Q_l(\lambda_c^{\text{U}})$ (energy released). Therefore, for this structure, the latent-heat is expressed as
\begin{equation}
\scalefont{0.85}
\begin{split}
 &Q_l(\lambda_c^{X}) = \\ 
 &\left(\frac{\delta -\lambda  (\lambda^* +\langle k \rangle) +\sqrt{(\delta +\lambda  (\lambda^* -\langle k \rangle))^2+4 \lambda^*  \langle k \rangle \lambda^2}}{2 \langle k \rangle \lambda } \right)_{\lambda = \lambda_c^X},
\end{split}
\end{equation}
where $\lambda_c^{X}$ can be ($\lambda_c^{\text{L}}$ or $\lambda_c^{\text{U}}$). In fact, this expression is true for any value of $\lambda$, but its physical interpretation is valid only near the discontinuity, which in turn depends on $\rho_c = \Theta^*$. We refer the reader to the SM for more details. 

Figure~\ref{fig:schematic} shows the general phenomenology of the system obtained from the analytical solution $-$i.e., the first order approximation$-$ of the equations describing the contagion dynamics for the hyperblob. As can be seen in Fig.~\ref{fig:schematic} (a), there are two possible solutions, $\rho^{\text{Lower}}$ and $\rho^{\text{Upper}}$. The solution depends on the initial conditions and the threshold, $\Theta^*$, which, together with the structure, defines a value $\rho_c$ where the dynamics exhibits a discontinuity. If $\rho(t=0) \geq \rho_c$, the solution is given by $\rho = \rho^{\text{Upper}}$ (forward phase diagram). On the other hand, if $\rho(t=0) < \rho_c$ and $\rho(t=0) \neq 0$, then $\rho = \rho^{\text{Lower}}$ (backward phase diagram). The up or down arrows show these solutions and directions as well as the size of the jump (i.e., the magnitude of the latent heat). Note, additionally, that the lower solution can exhibit a second-order phase transition. In Fig.~\ref{fig:schematic} (b) we instead represent the corresponding parameter space, which is composed of five distinct regions as explained in the figure caption. We assumed the most general case, where the lower solution has a transition from the absorbing state to an active state, here at $\lambda_c$. Note that, depending on the structure, the lower solution might have a vanishing critical point, i.e., $\lambda_c \rightarrow 0$, thus slightly changing this picture. We have also calculated, both analytically and numerically, the latent heat for this hypergraph structure. The results show that the absolute error between analytical and numerical simulations is of order $10^{-2} \sim 10^{-3}$ in hypergraphs with $N = 10^4$ (see SM section~\ref{sec:QS_regular} and Table~\ref{tab:latent_heat}), which indicates that the first-order approximation is accurate. 

In addition to the analysis of simple topologies, we also numerically verified this rich phenomenology on more heterogeneous structures. We confirmed that the order parameter can have two solutions, depending on the initial condition and the thresholds for hyperedges. Thus, generically, the solutions for a social contagion dynamics on hypergraphs can be expressed, mathematically, as 
\begin{eqnarray} 
&\rho^{\bigstar} =& 
 \begin{cases} \label{eq:rho_forward_gen}
  \rho^{\text{Lower}} \hspace{1cm} \text{if} \hspace{2mm} \rho^{\text{Lower}} < \rho_c\\
  \rho^{\text{Upper}} \hspace{1cm} \text{if} \hspace{2mm} \rho^{\text{Lower}} \geq \rho_c
 \end{cases} \\
&\rho^{*} =& 
 \begin{cases} \label{eq:rho_backward_gen}
  \rho^{\text{Upper}} \hspace{1cm} \text{if} \hspace{2mm} \rho^{\text{Upper}} \geq \rho_c\\
  \rho^{\text{Lower}} \hspace{1cm} \text{if} \hspace{2mm} \rho^{\text{Upper}} < \rho_c\\
 \end{cases} 
\end{eqnarray}
where $\rho^{\bigstar}$ is obtained if $\rho(t=0) < \rho_c$, and $\rho^{*}$ if $\rho(t=0) \geq \rho_c$, where $\rho_c$ is a global critical-mass, i.e., the value of $\rho$ at which the discontinuity appears. As before, we note that the lower solution (branch) might also exhibit a second-order (continuous) phase transition $-$denoted by $\lambda_c$ in Fig.~\ref{fig:schematic}$-$ from the absorbing state ($\rho = 0$) to the active state ($\rho > 0$). Furthermore, for a given hypergraph with fixed $\delta$ and $\lambda^*$, the discontinuity points are formally defined as
\begin{eqnarray} \label{eq:discontinuities}
 &\lambda_c^{\text{L}} =& \text{arg}_{\lambda} \left( \rho^{\text{Lower}}(\lambda, \delta, \lambda^*, N) = \rho_c \right) \\
 &\lambda_c^{\text{U}} =& \text{arg}_{\lambda} \left( \rho^{\text{Upper}}(\lambda, \delta, \lambda^*, N) = \rho_c \right),
\end{eqnarray}
thus, also defining the bi-stable region, $(\lambda_c^{\text{U}}, \lambda_c^{\text{L}})$.

\begin{figure}[t]
\includegraphics[width=\linewidth]{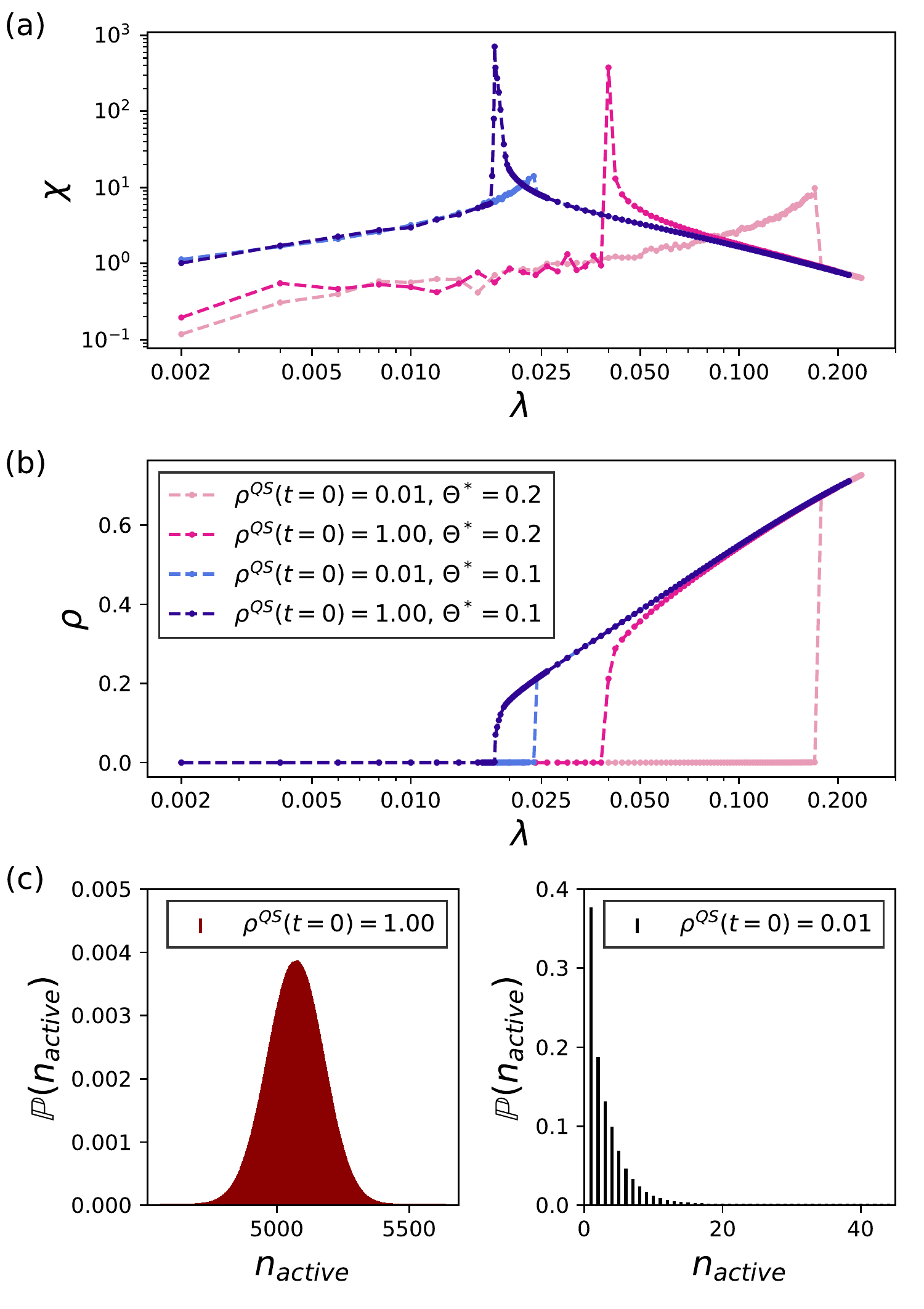}
\caption{Estimation of $\rho$ and $\chi$ using the QS method in a hypergraph with an exponential distribution of hyperedge cardinalities and $N = 10^4$. The dynamical parameter are: $\delta = 1.0$, $\lambda^* = \log_2(|e_j|)$ and $\Theta^* = 0.1, 0.2$. In (a) we present the the susceptibility, in (b) the order parameter. We considered two initial conditions for the QS method, $\rho^{QS}(t=0) = 0.01$, darker colors, and $\rho^{QS}(t=0) = 1.00$ lighter colors. In (c) the distribution of active node estimated using the QS method at $\lambda = 0.086$ and $\Theta^* = 0.2$ (the crossing between the two susceptibility curves, in Fig.~\ref{fig:HyperExp_QS} (a)).}
\label{fig:HyperExp_QS}
\end{figure}

Although a closed-solution for the general case is not possible, Monte Carlo simulations and numerical evaluation of Eq.~\ref{eq:first_order} are reasonable alternatives to characterize our system (see SM, section~\ref{sec:MC}).  Here, we focus on a hypergraph with an exponential distribution of cardinalities, $\Prob{|e|} \sim \mu \exp \left( -\mu |e| \right)$ with the constraint that $|e| \geq 2$. Dynamically, we set $\lambda^* = \log_2(|e|)$. This choice is arbitrary, but we choose here the $\log_2(|e|)$ function because it grows sublinearly. Note that, if a hyperedge cardinality goes to infinity, the average spreading value tends to zero, i.e., $\lim_{|e| \rightarrow \infty} \frac{\log_2(|e|)}{|e|}$ = 0. The impact of such a function is yet unknown, and we left this analysis for future work. 

Fig.~\ref{fig:HyperExp_QS} (a) and (b) show that the order parameter and the susceptibility follow the patterns expected for a first-order transition, i.e., both are discontinuous. Moreover, the order parameter is bi-stable, implying the presence of a hysteresis loop. This phenomenon is opposed to an SIS on a graph, where a second-order phase transition is characterized by a continuous behavior of the order parameter and a diverging susceptibility in the thermodynamic limit. Complementarily, Fig.~\ref{fig:HyperExp_QS} (c) shows the distribution of active nodes in the upper and lower branches. In the former, we have a bell-shaped distribution, similar to the super-critical regime of an SIS process~\cite{Ferreira2012, Arruda2018}. In the latter, we have a distribution peaked at one, similar to the subcritical regime (absorbing state) of an SIS process~\cite{Ferreira2012, Arruda2018}. We emphasize that Fig.~\ref{fig:HyperExp_QS} (c) displays the distribution of active nodes for the upper (left panel) and lower (right panel) branches and that the complete distribution for a given $\lambda$ in a region where both solutions exist is bimodal. It is also worth mentioning that, intuitively, one would expect lower cardinality hyperedges to be responsible for the lower branch, because they are easier to activate than the higher cardinality ones. Although $\Prob{|e| = 2} \geq \Prob{|e| = k}$, $k = 3, 4 ... N$ (exponential distribution), the largest connected component is very small, six nodes in the simulated hypergraph. This contrasts with regular cases such as the hyperblob (see SM, sections~\ref{sec:hyperblob} and~\ref{sec:Hyperstar}), in which the giant pairwise component has $N$ nodes. Furthermore, the generality of the reported phenomenological behavior suggests that the observed dynamics is a consequence of group-group interactions. This is further corroborated by additional results (reported in the SM) for a hypergraph with a power-law distribution of cardinality. In all systems, we found similar qualitative behavior for finite networks. 

Our results are important because they provide a theoretical foundation for, and a phenomenological explanation to, seemingly different experimental findings~\cite{Kanter1977, Drude1988, Grey2006, Centola2018}. These works reported critical mass levels needed to change an established equilibrium of $10 \%$ in some experiments and $30-40 \%$ in others, in apparent contradiction. Interestingly, in all of these experiments, individuals have group interactions instead of pairwise ones. The formalism here developed naturally brings forth plausible hypotheses for these observations and show that both ranges are possible. On the one hand, studies based on a single group suggest a threshold between $30\% \sim 40\%$, a situation that can be modeled as a single hyperedge in our formalism. On the other hand, a critical mass of $10 \%$ would correspond to a population that is composed by groups of diverse sizes, each one with a (larger) activation threshold. In other words, it is possible to have individual groups exhibiting a threshold $\Theta^*$ between $30\% \sim 40\%$, and at the same time a global critical mass, $\rho_c$, for the whole population of about $10 \%$ due to group intersections and interactions. A second reason that could explain the experimental findings is even simpler: admittedly, the fact that our model shows bi-stability also enables, for a given $\lambda$, two possible solutions for $\rho$ corresponding to the lower and the upper branches. That is, the system might be operating in the region where both solutions are larger than zero and stable.

In summary, in this paper we have developed a framework that allows to extend the study of social contagion models when group interactions are relevant. This is achieved by considering hypergraphs as the substrates that capture such many-to-many interactions. Several findings support the relevance of this methodology. First, our work opens the path to deal with new dynamical processes on top of higher-order models, and specifically on hypergraphs with no significant constraints. Secondly, we showed that simple dynamical processes can exhibit very rich dynamics, with different transitions, bistability and hysteresis. Ultimately, the uncovered phenomenology allows to explain seemingly contradictory experimental findings in which group interactions play a major role. Finally, we also mention that there are many interesting questions that arise from our work. For instance, if one assumes that energy is proportional to $\rho$, our model might display phenomena reminiscent of a Carnot cycle for social contexts, which might help to understand abrupt changes and oscillatory patterns in social behaviors. 


\begin{acknowledgments} GFA thanks E. Artiges, H. F. de Arruda, J. P. Rodriguez, L. Gallo and T. Peron for fruitful and inspiring discussions. GP acknowledges support from Compagnia San Paolo (ADnD project). YM acknowledges partial support from the Government of Aragon, Spain through grant E36-17R (FENOL), and by MINECO and FEDER funds (FIS2017-87519-P). GFA, GP, and YM acknowledge support from Intesa Sanpaolo Innovation Center. Research carried out using the computational resources of the Center for Mathematical Sciences Applied to Industry (CeMEAI) funded by FAPESP (grant 2013/07375-0). The funders had no role in study design, data collection, and analysis, decision to publish, or preparation of the manuscript. 
\end{acknowledgments}


\appendix
\section{Hypergraph structure} \label{sec:structure}

A hypergraph is formally defined as a set of nodes, $\mathcal{V} = \{ v_i \}$, where $N = |\mathcal{V}|$ is the number of nodes and a set of hyperedges $\mathcal{E} = \{ e_j \}$, where $e_j$ is a subset of $\mathcal{V}$ with arbitrary cardinality $|e_j|$. If $\max \left( |e_j| \right) = 2$ we recover a graph. On the other hand, if for each hyperedge with $|e_j| > 2$ its subsets are also contained in $\mathcal{E}$, we recover a simplicial complex. Fig.~\ref{fig:hypergraph_ex} (in the main text) illustrate a general hypergraph. In the same figure, we also show the comparison between the hypergraph and a graph simplification. In such simplification, we project a hyperedge as the set of all possible edges. In other words, a clique. This figure emphasizes the differences between both representations. Aside from the social examples presented in the main text, another interesting application of a hypergraph would be the collaboration hypergraph (as opposed to the collaboration network). In the network case, if two authors publish a paper together, they share a link. However, from the network, it is impossible to recover the papers. Interestingly, modeling papers as hyperedges would allow complete information encoding. We remark that this example is not explored in this paper, but left as a future application of our work.

In order to obtain analytical insights about our dynamics, we use hypergraphs with structural symmetry. Here we focus on two regular cases: (i) star hypergraph, here called hyperstar, and (ii) hyperblob, a random regular hypergraph. The first is defined as a central node connected to all other nodes with hyperedges with cardinality two, i.e., simple edges forming a star graph, and a hyperedge containing all the nodes, thus having cardinality $N$. The second case is defined as a random regular network as the hyperedges with cardinality two and a hyperedge with all nodes, thus with cardinality $N$.

Aside from the regular symmetric structures, we also use more general structures. Recently, in~\cite{Chodrow2019} the author proposed a configuration model for hypergraphs. Since we are focusing our analysis on the dynamics, here we follow a relatively simpler approach. First of all, we define the number of hyperedges, $M$. Then, we sample the cardinality of this hyperedges from an arbitrary distribution. In other words, we are interested in the analysis of the heterogeneity in $|e_j|$, given by $\Prob{|e|}$. As a constrain, we impose that $2 \leq |e_j| \leq N$, since if $|e_j| \leq 2$ the node is isolated by pairwise interactions. Next, we construct each hyperedge by uniformly sampling nodes from $\mathcal{V}$. To evaluate different levels of structural heterogeneity, in terms of $\Prob{|e|}$, we follow two distribution, the exponential, formally given as
\begin{equation}
 \Prob{|e|} \sim 
 \begin{cases}
  \mu \exp \left( -\mu |e| \right) \hspace{3mm} &\text{if} \hspace{2mm} |e| \geq 2\\
  0 &\text{otherwise}
 \end{cases}
\end{equation}
where $\mu$ defines the average and variance and a power-law, defined as 
\begin{equation}
 \Prob{|e|} \sim 
 \begin{cases}
  |e|^{-\gamma} \hspace{3mm} &\text{if} \hspace{2mm} |e| \geq 2\\
  0 &\text{otherwise}
 \end{cases}
\end{equation}
where $\gamma$ control the heterogeneity. In both cases, the highest probability is obtained at $\Prob{|e|=2}$. This is a reasonable assumption since it is already known in the literature that complex networks are able to represent many real systems~\cite{Arruda2018}. The exponential case models a structure where the groups are close to an average value. On the other hand, for the power-law case we expect a larger variance, depending on $\gamma$. It might even diverge if $\gamma < 3$. Structurally, this implies that we can have hyperedges with all the elements in $\mathcal{V}$. Here, we used $\gamma = 2.25$.

\section{Analysis of hyperedge activity} \label{sec:F}

The probability that a given hyperedge, $e_j$, is active is given as
\begin{equation}
F(\Theta) = \sum_{k = \Theta_j}^{|e_j|} \mathbb{P}_{n} \left(K=k \right),
\end{equation}
where $\mathbb{P}_{n}$ is given by Eq.~\ref{eq:prob_pb} or Eq.~\ref{eq:dft_pn}. In the following sections we explore: (i) the cases where we have structural symmetries, thus a Bernoulli distribution and (ii) the general case.

\subsection{Bernoulli distribution: induced by structural symmetries}

Due to symmetries, the Poisson binomial distribution reduces to a binomial distribution. This reduction applies to star and the homogeneous hypergraphs cases. In this cases, the functions $F(\Theta)$ will depend only on the probability that a leaf is active, $y_l$, and the number of leaves, $(N-1)$, in the star case. Conversely, in the homogeneous case, it will depend on the individual probability, $y$, and the number of individuals, $N$. Furthermore, observe that in the thermodynamic limit, we assume that $\Theta = \ceil[\big]{\Theta^* N}$, where $\Theta^*$ is a real number representing the fraction of active nodes.

\begin{figure}[t]
\includegraphics[width=\linewidth]{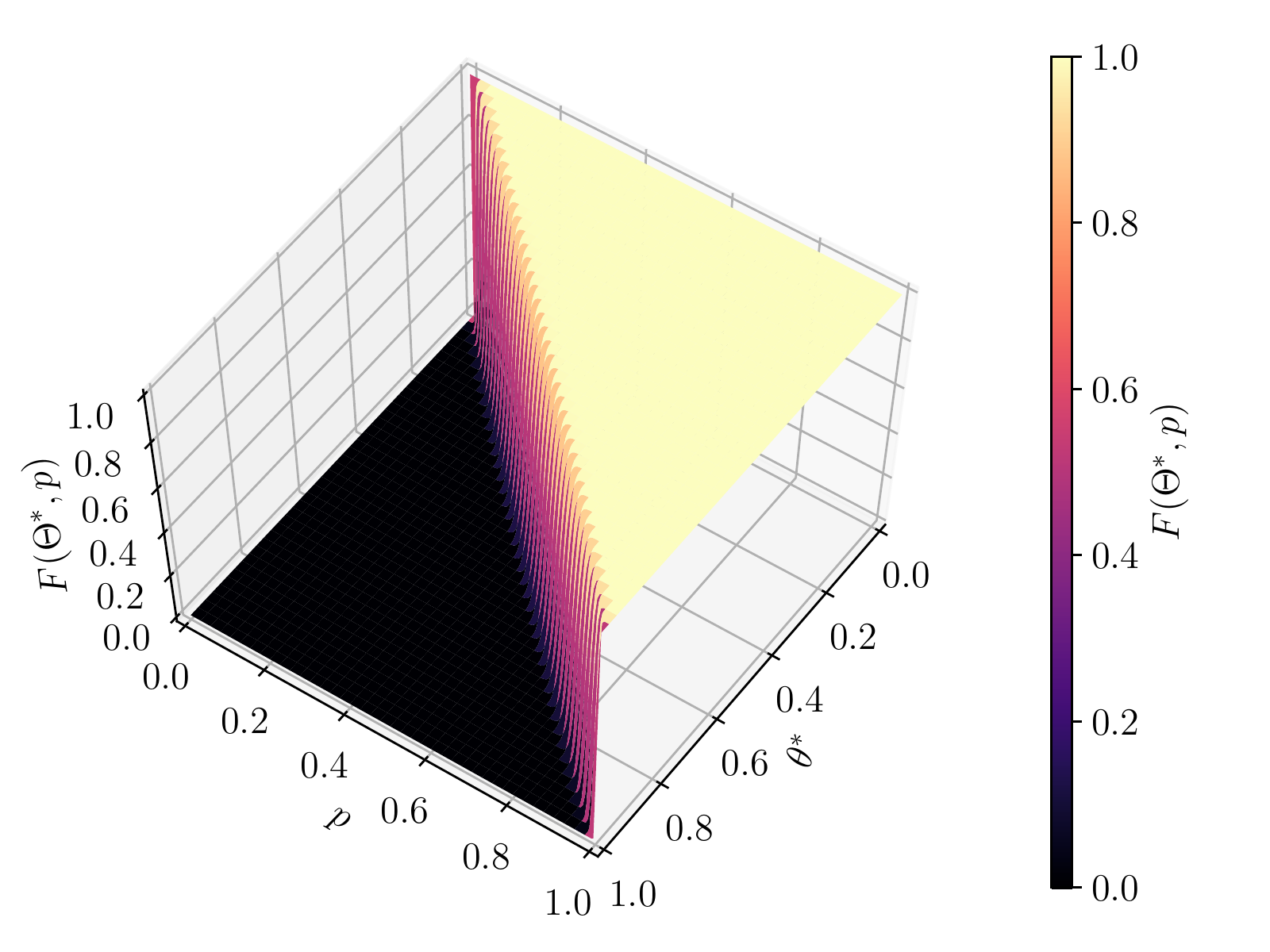}
\caption{Function $F(\Theta^*, p)$ numerically evaluated using equations~\ref{eq:F_0} and~\ref{eq:Pn_star} $N=10^3$. The finite size effects are still visible, but it is very subtle and its tendency is clear.}
\label{fig:F}
\end{figure}

\begin{figure}[t]
\includegraphics[width=\linewidth]{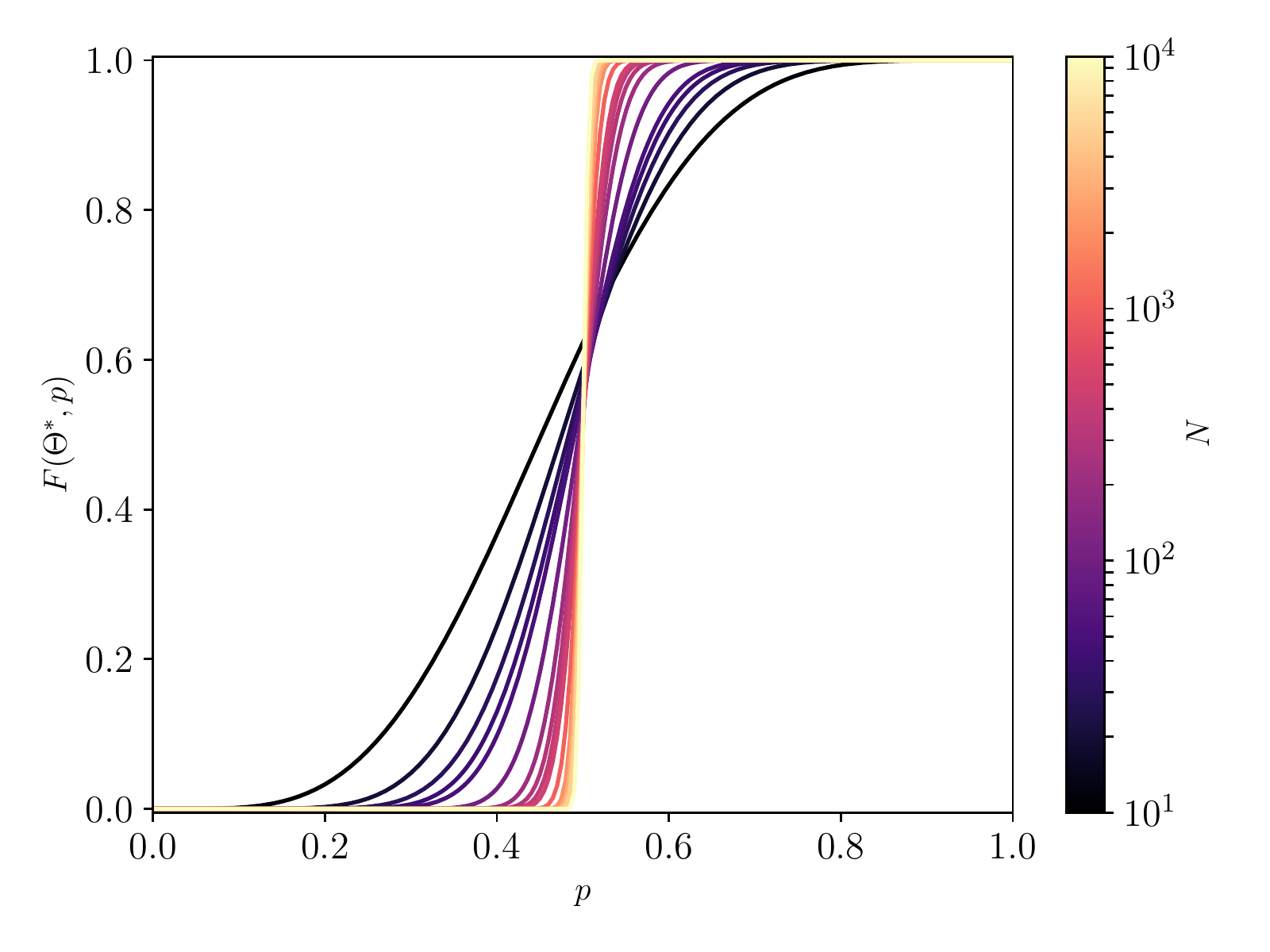}
\caption{Function $F(\Theta^*, p)$ numerically evaluated for different values of $p$ and $\Theta^*=0.5$ using equations~\ref{eq:F_0} and changing $N$. This illustrates the convergence to Eq.~\ref{eq:F_infty} as $N$ increases.}
\label{fig:F_N}
\end{figure}

Next, from the hyperstar case, denoting $\tilde{Y} = \sum_{i=1}^{N-1} (Y_l)_i$ as the number of active nodes, we have, 
\begin{eqnarray}
 \E{\frac{\tilde{Y}}{N-1}} = p, \\
 \mathbb{V} \left( \frac{\tilde{Y}}{N-1} \right) = \frac{p(1-p)}{N-1},
\end{eqnarray}
where $p$ is the parameter of the binomial distribution (or probability of success in a trial). Evidently, the average value does not depend on $N$, while the variance tends to zero in the thermodynamic limit. Since such distribution is centered in its average values, we can conclude that
\begin{equation}
 \lim_{N \rightarrow \infty} \mathbb{P}_{N} \left(\tilde{Y}=\tilde{y} \right) = 
 \begin{cases}
  1, \hspace*{1cm} \text{if} \hspace*{2mm} \tilde{y} = p \\
  0, \hspace*{1cm} \text{otherwise}
 \end{cases}
\end{equation}
and consequently, 
\begin{equation} \label{eq:F_infty}
 F(\Theta^*, p) = 
 \begin{cases}
  1, \hspace*{1cm} \text{if} \hspace*{2mm} \Theta^* \geq p \\
  0, \hspace*{1cm} \text{otherwise}
 \end{cases}.
\end{equation}
Fig.~\ref{fig:F} shows an example of $F(\Theta^*, p)$ for a range of parameters and a finite $N = 10^3$. Although the it is a finite number of nodes, the trend is clear. Furthermore, the convergence to Eq.~\ref{eq:F_infty} is illustrated in Fig~\ref{fig:F_N}, where $F(\Theta^*, p)$ is evaluated for a range of $p$ and $\Theta^* = 0.5$ for different values on $N$.

Dynamically, $p = y_l(t)$ in the star case or $p = y(t)$ in the homogeneous case, which are a function of time. When the fraction of active nodes reaches the threshold $\Theta^*$, the functions $F^{Y_c=0}$, $F^{Y_c=Y_l=0}$ and $F^{Y_c=1, Y_l=0}$ (defined in the next sections) will be close to one, ``activating'' the group spreading. If the spreading of the group is sufficiently fast, the spread will be effective, and $p$ will also increase. At this moment, there will be a competition of spreading processes (the standard contact and the group spreading) against the annihilation mechanism. On the other hand, if the fraction of active nodes does not reach the threshold $\Theta^*$, then only the standard contact takes place, and it competes alone with the annihilation mechanism.

\subsection{Poisson binomial distribution: the general case}

In the general case, nodes will have different patterns of connections and the hyperedge cardinalities can be arbitrarily distributed. Thus, there is no symmetries and the probabilities $y_i$ will follow an unknown distribution. In this scenario, for each hyperedge with cardinality $n$, the average value and variance can be expressed as
\begin{eqnarray}
 \E{\frac{\tilde{Y}}{n-1}} = \frac{\sum_i p_i}{n-1}, \\
 \mathbb{V} \left( \frac{\tilde{Y}}{n-1} \right) = \frac{\sum_i p_i(1-p_i)}{(n-1)^2},
\end{eqnarray}
where $p_i = y_i(t)$ in this context. Note that, for sufficiently large hyperedges, the variance is also expected to vanish, suggesting that an approximation for $F(\Theta^*, \{p_i\})$ might not be useful. 

We also remark that in the most general cases, hyperedges might have relatively small cardinalities, does not allow us to use this approximation. Indeed, for the simplicial complex model in~\cite{Iacopo2019} (only triangles) our approximation is not valid.

\section{Analysis of a homogeneous hypergraph} \label{sec:hyperblob}

\subsection{Definition}

In this section we use the homogeneous hypergraph described in section~\ref{sec:structure}. This structure have a strong symmetry, allowing us to reduce Eq.~\ref{eq:first_order} to a single equation. Therefore, we have
\begin{equation} \label{eq:rrn}
 \dfrac{d y}{dt} = -\delta y + \lambda (1 - y) \left[ \langle k \rangle y + \lambda^*(N) F^{Y_i=0} \right],
\end{equation}
where
\begin{equation}
 F^{Y_i=0} \left( \Theta \right) = 1 - \sum_{k=0}^{\Theta-1} \mathbb{P}_{N-1} \left(K=k \right).
\end{equation}
Note that this is not a quenched formalism since this equation describes a family of structures ranging from lattices to random regular networks. Besides, one might also expect that if the hyperedges with cardinality $2$ are defined as an Erd\H{o}s-R\'enyi network, this equation would also capture its qualitative behavior.

\subsection{Steady-state analysis}

Considering the steady state, i.e. $\dfrac{d y}{dt} = 0$. Furthermore, lets assume that $N$ is sufficiently large, but finite. Therefore, $F^{Y_c=0} \approx F^{Y_c=Y_l=0} \approx F^{Y_c=1, Y_l=0} \approx F(\Theta, p)$. Note that as $N$ increases the latter assumption also improves. Thus, we have
\begin{equation} \label{eq:rrn_ss}
 0 \approx -\delta y + \lambda (1 - y) \left[ \langle k \rangle y + \lambda^*(N) F(\Theta^*,p) \right] = f(y),
\end{equation}

\begin{figure*}[t]
\includegraphics[width=\linewidth]{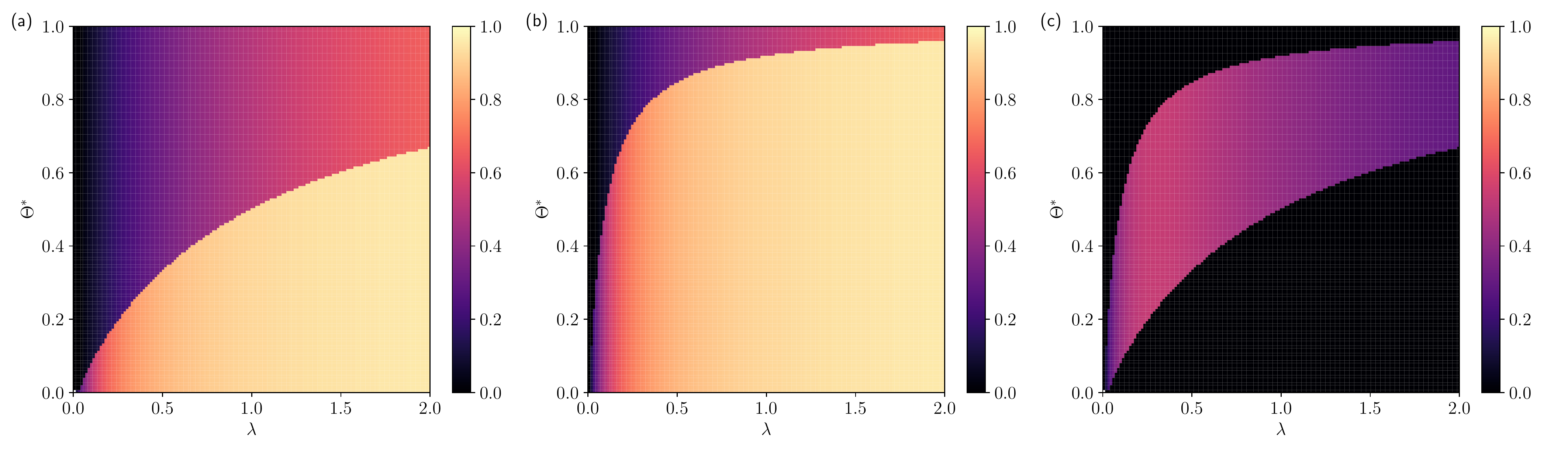}
\caption{Phase diagram for the RRN hypegraph with $k = 5$, $\delta = 1$ and $\lambda^*(|e_j|) = \log_2(|e_j|)$. In (a)-(c) the colormaps are obtained changing $\lambda$ and $\Theta^*$. In (a) the solution of Eq.~\ref{eq:rho_forward}, in (b) the solution of Eq.~\ref{eq:rho_backward} and in (c) the latent heat (i.e., difference between (b) and (a)), emphasizing the hysteresis loop.}
\label{fig:analytical_pd_rrn}
\end{figure*}

If $F(\Theta, p) = 0$ or $\lambda^* = 0$ we recover the QMF solutions for the SIS in a random regular graph (RRN). Formally,
\begin{equation}
\rho^{\text{Lower}} = 
 \begin{cases}
  1 - \frac{\delta}{\langle k \rangle \lambda}, \hspace*{1cm} \text{if} \hspace*{2mm} \frac{\lambda}{\delta} \geq \frac{1}{\langle k \rangle}\\
  0, \hspace*{2.1cm} \text{otherwise}
 \end{cases}.
\end{equation}

On the other hand, if we assume that we are in the regime where $F(\Theta^*,p) = 1$, Eq.~\ref{eq:rrn_ss} can be solved as
{ \scalefont{0.95}
\begin{equation} \label{eq:y_upper_rrn}
 y^{\pm} = \frac{-\delta + \langle k \rangle \lambda - \lambda^* \lambda \pm \sqrt{ 4 \langle k \rangle \lambda^* \lambda^2 + (\delta + (-\langle k \rangle + \lambda^*) \lambda)^2}}{(2 \langle k \rangle \lambda)}
\end{equation}
}
where $y^{+}$ is the feasible solution. Note that $y^{-}$ might lead to negative values. Therefore, in the regime $F(\Theta^*,p) = 1$, the order parameter can be expressed as
{ \scalefont{0.8}
\begin{equation}
 \rho^{\text{Upper}} = \frac{-\delta + \langle k \rangle \lambda - \lambda^* \lambda + \sqrt{ 4 \langle k \rangle \lambda^* \lambda^2 + (\delta + (-\langle k \rangle + \lambda^*) \lambda)^2}}{(2 \langle k \rangle \lambda)}.
\end{equation}
}

Thus, from both solutions, we can obtain the order parameter as
\begin{eqnarray} 
&\rho^{\bigstar} =& 
 \begin{cases} \label{eq:rho_rrn_forward}
  \rho^{\text{Lower}} \hspace{1cm} \text{if} \hspace{2mm} \rho^{\text{Lower}} < \Theta^*\\
  \rho^{\text{Upper}} \hspace{1cm} \text{otherwise}
 \end{cases} \\
&\rho^{*} =& 
 \begin{cases} \label{eq:rho_rrn_backward}
  \rho^{\text{Upper}} \hspace{1cm} \text{if} \hspace{2mm} \rho^{\text{Upper}} \geq \Theta^*\\
  \rho^{\text{Lower}} \hspace{1cm} \text{otherwise},
 \end{cases}.
\end{eqnarray}
Note that our approximation is based in two ideas: (i) we assumed that the system is sufficiently large, so the function $F(\Theta^*,p)$ is a good approximation for $F^{Y_i=0}$ and (ii) the state of the individuals is independent.

Figure~\ref{fig:analytical_pd_rrn} shows the the solutions of equations~\ref{eq:rho_rrn_forward} and~\ref{eq:rho_rrn_backward} for different parameters $(\lambda, \Theta^*)$.

\subsection{Local stability analysis} \label{sec:stability_hyperblob}

The solutions $\rho^{\text{Lower}}$ and $\rho^{\text{Upper}}$ are fixed points of the ODE system of Eq.~\ref{eq:rrn}. However, we should prove that those solutions are also stable. To do so, we have to calculate the derivative of $f(y)$ and evaluate it near the fixed point. Therefore, for the first solution, $\rho^{\text{Lower}}$, we have that
\begin{equation} \label{eq:dfdy_rrn}
 \left(\dfrac{d f(y)}{dy}\right)_{y=\rho^{\bigstar}} = 
 \begin{cases}
  \delta -\langle k \rangle \lambda , \hspace*{1cm} \text{if} \hspace*{2mm} \frac{\lambda}{\delta} \geq \frac{1}{\langle k \rangle}\\
  \langle k \rangle \lambda -\delta, \hspace*{1cm} \text{otherwise}
 \end{cases},
\end{equation}
which is stable (negative) for any parameter above the critical point $\frac{\lambda}{\delta} \geq \frac{1}{\langle k \rangle}$. Furthermore, the second line of~\ref{eq:dfdy_rrn}, signs the bifurcation that accounts for the second order phase transition in the classical model, i.e. in a graph.

Next, considering the regime where $F(\Theta^*,p) = 1$, the derivative can expressed as
\begin{equation}
 \begin{split}
 \left(\dfrac{d f(y)}{dy}\right)_{y=y^{+}} = - \sqrt{ 4 \langle k \rangle \lambda^* \lambda^2 + (\delta + (-\langle k \rangle + \lambda^*) \lambda)^2},
 \end{split}
\end{equation}
which is negative for any set of parameters $\lambda >0$, $\delta>0$, $\lambda^*>0$ and $\langle k \rangle > 0$. Thus, it is also an stable solution. Moreover, note that $y^{-}$ in Eq.~\ref{eq:y_upper_rrn} would lead to an unstable solution.

With this analysis we show that the equations~\ref{eq:rho_rrn_forward} and~\ref{eq:rho_rrn_backward} are stable, therefore there is a bi-stable region. Physically, this means that, depending on the initial condition, the process will have one of these solutions. It is worth mentioning that, for finite size systems, a fluctuation can induce the switch between solutions. However, the likelihood of this happening decreases as the system size increases, vanishing in the thermodynamic limit.

\subsection{Critical values}

The critical values are obtained as the points when $\Theta^*$ crosses $\rho^{\text{Lower}}$ and $\rho^{\text{Upper}}$ for the lower and upper solution respectively. Formally,
\begin{eqnarray}
 &\lambda_c^{\text{L}} =& \text{arg}_{\lambda} \left( \rho^{\text{Lower}}(\lambda, \delta, \lambda^*, N) = \Theta^* \right) \\
 &\lambda_c^{\text{U}} =& \text{arg}_{\lambda} \left( \rho^{\text{Upper}}(\lambda, \delta, \lambda^*, N) = \Theta^* \right),
\end{eqnarray}
where, we are assuming that $\delta$, $\lambda^*$, $N$ and $\langle k \rangle$ are kept fixed and the critical point is calculated in terms of $\lambda$. Note that more general expressions might be calculated using the same principle.

From equations~\ref{eq:rho_rrn_forward} and~\ref{eq:rho_rrn_backward}, it is clear that the critical points appear when the lower and upper solution crosses the threshold $\Theta^*$. Since we have closed expressions for the order parameter as a function of the control parameters we can solve this equation. Thus, assuming that $\lambda$ is our control parameter and $\delta$, $\lambda^*$ and $\langle k \rangle$ are kept fixed, we have  
\begin{eqnarray}
 \lambda_c^{\text{L}} &=& \frac{\delta}{\langle k \rangle - \Theta^*  \langle k \rangle} \\
 \lambda_c^{\text{U}} &=& -\frac{\delta  \Theta^* }{\lambda^*  \Theta^* -\lambda^* +(\Theta^*)^2 \langle k \rangle - \Theta^* \langle k \rangle}.
\end{eqnarray}
Note that we were only able to obtain this relation due to the simplicity of or model.

\subsection{``Latent heat''}

Physically, the energy released or absorbed during a constant-temperature process is called latent heat. Here we can also define a similar concept. In complex systems in general the definition of energy itself is not trivial, however, in our context, it is reasonable to assume that it is proportional to the order parameter, $\rho$. Furthermore, the control parameter in many thermodynamic systems is the temperature. Here, the analogous would be the spreading rate $\lambda$. Note that, in this case, we are assuming that the other parameters of the model, $\delta$, $\lambda^*$, $\langle k \rangle$ and $\Theta^*$, are kept fixed. Thus, the analogous to the latent heat is the difference between upper and lower solution. Formally,
\begin{equation}
 Q_l(\lambda_c^{X}) = \left( \rho^{\text{Upper}}(\lambda, \delta, \lambda^*, N) - \rho^{\text{Lower}}(\lambda, \delta, \lambda^*, N) \right)_{\lambda = \lambda_c^X}
\end{equation}
where $Q_l(\lambda_c^{X})$ can be $Q_l(\lambda_c^{\text{L}})$ or $Q_l(\lambda_c^{\text{U}})$. Furthermore, note that the concept of latent heat is intrinsically connected with first-order phase transitions.

Thus, in the homogeneous case, the latent heat can be expressed analytically as
\begin{equation}
\scalefont{0.85}
\begin{split}
 &Q_l(\lambda_c^{X}) = \\ 
 &\left(\frac{\delta -\lambda  (\lambda^* +\langle k \rangle) +\sqrt{(\delta +\lambda  (\lambda^* -\langle k \rangle))^2+4 \lambda^*  \langle k \rangle \lambda^2}}{2 \langle k \rangle \lambda } \right)_{\lambda = \lambda_c^X},
\end{split}
\end{equation}
where $\lambda_c^{X}$ can be $(\lambda_c^{\text{L}}$ or $\lambda_c^{\text{U}}$. In fact, this expression is true for any value of $\lambda$, but its physical interpretation is valid only at the discontinuity, which is, in its turn, dependent on the parameter $\Theta^*$.

In terms of a social process, the latent heat interpretation is the fraction of individuals we have to add or remove to move the dynamics from one solution to the other.

\section{Analysis of the Hyperstar} \label{sec:Hyperstar}

\subsection{Definition}

In order to obtain some analytical insights from our approximation we consider a hypergraph composed by a star graph and single hyperedge with all the nodes. Due to the symmetries of this configuration, Eq.~\ref{eq:first_order} reduces to
\begin{equation} \label{eq:star}
 \begin{cases}
 \!\begin{aligned}
  \dfrac{d y_c}{dt} &= -\delta y_c + \lambda (1 - y_c) \left[ (N-1)y_l + \lambda^*(N) F^{Y_c=0} \right]\\
  \dfrac{d y_l}{dt} &= -\delta y_l + \lambda (1 - y_l) \times \\
  &\times \left[y_c + \lambda^*(N) \left( (1 - y_c) F^{Y_c=Y_l=0} + y_c  F^{Y_c=1, Y_l=0} \right) \right],
  \end{aligned}
 \end{cases}
\end{equation}
where $y_c$ and $y_l$ are respectively the probability that the central node and a leaf are active and the spreading probabilities are given as
\begin{eqnarray}
 &F^{Y_c=0} \left( \Theta \right) = 1 - \sum_{k=0}^{\Theta-1} \mathbb{P}_{N-1} \left(K=k \right), \label{eq:F_0}\\
 &F^{Y_c=Y_l=0} \left( \Theta \right) = 1 - \sum_{k=0}^{\Theta-1} \mathbb{P}_{N-2} \left(K=k \right), \\
 &F^{Y_c=1, Y_l=0} \left( \Theta \right) = 1 - \sum_{k=0}^{\Theta-2} \mathbb{P}_{N-2} \left(K=k \right),
\end{eqnarray}
where the dependency in $\Theta$ was suppressed in Eq.~\ref{eq:star} and the superscript $Y_l = 0$ on the $F$ functions indicates that one of the leafs is considered to be inactive. Furthermore, 
\begin{equation} \label{eq:Pn_star}
  \mathbb{P}_{n} \left(K=k \right) = \frac{1}{n+1} \sum\limits_{l=0}^n C^{-lk} \left( 1+(C^l-1) y_l \right)^n.
\end{equation}

In the star hypergraph, the order parameter reduces to
\begin{equation}
 \rho = \frac{1}{N} \left( y_c + (N-1)y_l \right),
\end{equation}
and, in the steady-state, $\dfrac{d y_c}{dt} = \dfrac{d y_l}{dt} = 0$.

\subsection{Steady-state analysis}

Considering the steady state, i.e. $\dfrac{d y_c}{dt} = \dfrac{d y_l}{dt} = 0$. Furthermore, lets assume that $N$ is sufficiently large, but finite. Therefore, $F^{Y_c=0} \approx F^{Y_c=Y_l=0} \approx F^{Y_c=1, Y_l=0} \approx F(\Theta, p)$. Note that as $N$ increases the latter assumption also improves. Thus, we have
\begin{equation} \label{eq:ss_star}
 \begin{cases}
 \!\begin{aligned}
  0 & \approx -\delta y_c + \lambda (1 - y_c) \left[ (N-1)y_l + \lambda^* F(\Theta^*) \right] \\
  0 & \approx -\delta y_l + \lambda (1 - y_l) \left[y_c + \lambda^* F(\Theta^*) \right].
  \end{aligned}
 \end{cases}
\end{equation}

If $F(\Theta, p) = 0$ or $\lambda^* = 0$ we recover the QMF solutions for the SIS in a star graph. Formally,
\begin{eqnarray}
&y_c =& \label{eq:yc_lower}
 \begin{cases} 
  \frac{-\delta ^2+(N-1) \lambda ^2}{\lambda  (\delta +(N-1) \lambda )} \hspace{1.3cm} &\text{if} \hspace{2mm}  \frac{\lambda}{\delta} > \frac{1}{\sqrt{N-1}} \\
  0 & \text{otherwise} 
 \end{cases}\\
&y_l =& \label{eq:yl_lower}
 \begin{cases}
  \frac{-\frac{\delta ^2}{N-1}+\lambda ^2}{\lambda  (\delta +\lambda )}  & \hspace{.3cm} \text{if} \hspace{2mm}  \frac{\lambda}{\delta} > \frac{1}{\sqrt{N-1}} \\
  0 & \hspace{.3cm} \text{otherwise}
 \end{cases},
\end{eqnarray}
where the critical point emerges naturally, i.e., $\frac{\lambda}{\delta} > \frac{1}{\sqrt{N-1}}$. For the sake of completeness, the order parameter is expressed as
\begin{equation} \label{eq:sis_qmf}
 \rho^{\text{Lower}}=
 \begin{cases}
  \frac{(2 \delta +N \lambda ) \left((N-1) \lambda ^2 - \delta^2\right)}{N \lambda  (\delta +\lambda ) (\delta +(N-1) \lambda )} \hspace{1.3cm} &\text{if} \hspace{2mm}  \frac{\lambda}{\delta} > \frac{1}{\sqrt{N-1}} \\
  0  &\text{otherwise} 
 \end{cases}.
\end{equation}
Note that this result only confirms the QMF theory for a finite star and that it is compatible with a second order phase transition. Besides, the approximations on the functions $F$ do not affect these results. In the thermodynamic limit,
\begin{equation} \label{eq:rho_inf}
 \lim_{N \rightarrow \infty} \rho^{\text{Lower}}=\frac{\lambda }{\delta +\lambda },
\end{equation}
where the critical point also goes to zero, $\lim_{N \rightarrow \infty} \frac{1}{\sqrt{N-1}} = 0$. Phenomenologically, this implies a vanishing critical point. This is in agreement with the mean-field predictions for an SIS in a star graph.

\begin{figure*}[t]
\includegraphics[width=\linewidth]{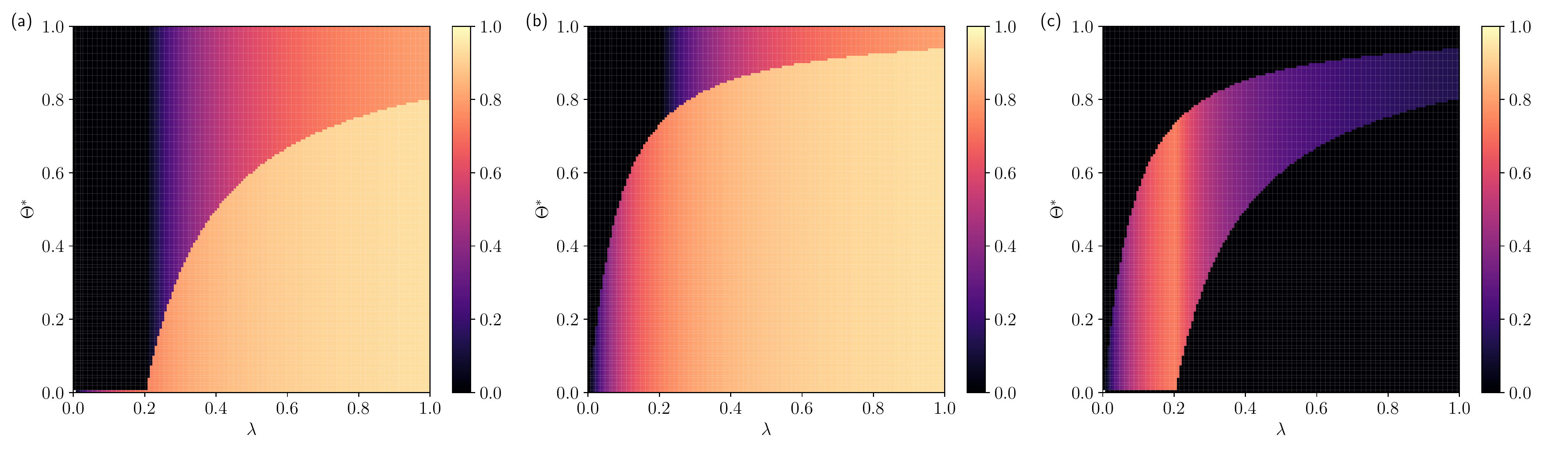}
\caption{Phase diagram for the hyperstar with $N = 10^3$, $\delta = 1$ and $\lambda^*(|e_j|) = \log_2(|e_j|)$. In (a)-(c) the colormaps are obtained changing $\lambda$ and $\Theta^*$. In (a) the solution of Eq.~\ref{eq:rho_forward}, in (b) the solution of Eq.~\ref{eq:rho_backward} and in (c) the latent heat (i.e., difference between (b) and (a)), emphasizing the hysteresis loop.}
\label{fig:analytical_pd}
\end{figure*}

Since we assumed that $N$ is sufficiently large and $F(\Theta, p)$ is a limited function, in the regime $F(\Theta, p) = 1$, we can solve Eq.~\ref{eq:ss_star} as
\begin{widetext}
  \begin{equation} \label{eq:ss_star_sol}
   \begin{cases}
        y_c^{\pm} = -\frac{\delta ^2+2 \delta  \lambda^*  \lambda +(\lambda^* -1) \lambda ^2 (\lambda^* + N-1) \pm    \sqrt{\left(\delta ^2+2 \delta  \lambda^*  \lambda +(\lambda^* +1) \lambda ^2 (\lambda^* + 1-N)\right)^2+4 \lambda^*  \lambda ^2 (N-1) (\delta +\lambda^*  \lambda +\lambda )^2}}{2 \lambda  (\delta +\lambda  (\lambda^* + N-1))} \\ 

        y_l^{\pm} = -\frac{\delta ^2+2 \delta  \lambda^*  \lambda - (\lambda^* +1) \lambda ^2 (\lambda^* + N-1) \pm \sqrt{\left(\delta ^2+2 \delta  \lambda^*  \lambda +(\lambda^* +1) \lambda ^2 (\lambda^* + 1-N)\right)^2+4 \lambda^*  \lambda ^2 (N-1) (\delta +\lambda^*  \lambda +\lambda )^2}}{2 \lambda  (N-1) (\delta +\lambda^*  \lambda +\lambda )} 
\end{cases},
 \end{equation}
\end{widetext}
where we have two solutions for each probability. Moreover, note that the ``$+$'' (``$-$'') solution of the first equation matches with the ``$+$'' (``$-$'') solution of the second. Next, assuming the thermodynamic limit in Eq.~\ref{eq:ss_star_sol} we have 
\begin{eqnarray}
&\lim_{N \rightarrow \infty} y_c =&
 \begin{cases}
  -\lambda^* \hspace{1cm} \text{unfeasible} \\
  1 
 \end{cases},\\
&\lim_{N \rightarrow \infty} y_l =&
 \begin{cases}
  0 \hspace{1.5cm} \text{unfeasible} \\
  \frac{(\lambda^* +1) \lambda }{\delta + \lambda^* \lambda +\lambda } 
 \end{cases}.
\end{eqnarray}
Note that the first solution, i.e. the solutions with the ``$+$'' signal, are unfeasible. It is clear that $y_l = 0$ implies that $y_c = -\lambda^*$, as can be verified in the second equation of Eq.~\ref{eq:ss_star}. Hence, in the thermodynamic limit, after the threshold $\Theta^*$, the order parameter is expressed as
\begin{equation} \label{eq:rho*_inf}
 \rho^{\text{Upper}} = \frac{(\lambda^* +1) \lambda }{\delta + (\lambda^* +1) \lambda }.
\end{equation}

Note that equations~\ref{eq:sis_qmf} and~\ref{eq:ss_star_sol} (only the ``$+$'' solution) provide a solution that depend on $N$, which should be large enough, but finite. On the other hand, equations~\ref{eq:rho_inf} and~\ref{eq:rho*_inf} provide its solution on the thermodynamic limit. Furthermore, with this analysis we showed that, for a sufficiently large $N$, the order parameter can be expressed as
\begin{eqnarray} 
&\rho^{\bigstar} =& 
 \begin{cases} \label{eq:rho_forward}
  \rho^{\text{Lower}} \hspace{1cm} \text{if} \hspace{2mm} \rho^{\text{Lower}} < \Theta^*\\
  \rho^{\text{Upper}} \hspace{1cm} \text{otherwise}
 \end{cases} \\
&\rho^{*} =& 
 \begin{cases} \label{eq:rho_backward}
  \rho^{\text{Upper}} \hspace{1cm} \text{if} \hspace{2mm} \rho^{\text{Upper}} \geq \Theta^*\\
  \rho^{\text{Lower}} \hspace{1cm} \text{otherwise},
 \end{cases} 
\end{eqnarray}
where $\rho^{\bigstar}$ is obtained if $\rho(t=0) < \Theta^*$ and $\rho^{*}$ is obtained if $\rho(t=0) \geq \Theta^*$. Phenomenologically, this means that we have two possible solutions, $\rho^{\bigstar}$ and $\rho^{*}$, possibly implying in a hysteresis loop due to a bi-stable region. The first solution is the same as for the SIS in a star graph, while the second is the result of the large hyperedge activation. 

Figure~\ref{fig:analytical_pd} shows the the solutions of equations~\ref{eq:rho_forward} and~\ref{eq:rho_backward} for different parameters $(\lambda, \Theta^*)$.

\subsection{Local stability analysis} \label{sec:stability_hyperstar}

To understand whether the $\rho$ solutions are stable we study the Jacobian of the system, following a local stability analysis. Here we are also assuming a large but finite system. As a consequence, the $F$ functions can be approximated as $F(\Theta^*, p)$. Note that $F(\Theta^*, p)$ is not differentiable in the whole domain. In fact there is a discontinuity for $\Theta^* = p$. Thus its derivative is not defined in those points. Restricting ourselves to the part of the domain where the function is continuous, the Jacobian matrix can be expressed as
\begin{equation}
 \J = -\delta \I +
 \lambda \left( \J_1 + \J_2 \right),
\end{equation}
where
\begin{eqnarray}
&\J_1 =& \left(
 \begin{array}{cc}
-(N-1) y_l  & (N-1) (1-y_c)  \\
 (1-y_l)  & -y_c  \\
\end{array}
\right) \\
&\J_2 \approx& -\lambda^* F(\Theta^*,p) \I,
\end{eqnarray}
Note that $\J_2$ would not be a diagonal matrix if we had assumed its complete dependency with $y_l$. Besides, only $\J_2$ contains the discontinuous part. Next, in order to determine if a given solution is stable we must evaluate the eigenvalues of $\J$ for the solutions of $y_l$ and $y_c$ obtained in the previous section.

Our analysis has three different cases: (i) $y_c = y_l = 0$; (ii) $F(\Theta^*,p) = 0$, which also imply that $y_c$ and $y_l$ follow equations~\ref{eq:yc_lower} and~\ref{eq:yl_lower}, respectively; and (iii) $F(\Theta^*,p) = 1$, implying that $y_c$ and $y_l$ follow the ``$-$'' solutions of Eq.~\ref{eq:ss_star_sol}. Note that, due to our approximation on the $F$ functions, we either assume that we are in one regime or the other solution. In this way, we can extract meaningful results. The first case is trivial. Indeed it is the absorbing state of our dynamics. The second case corresponds to the lower solution, which is also the solution of an SIS on a star graph. In this scenario, both eigenvalues of $\J$ are negative if $\frac{\lambda}{\delta} > \frac{1}{\sqrt{N-1}}$. In fact, one might also use the first equation of equations~\ref{eq:yc_lower} and~\ref{eq:yl_lower} to derive the critical point. In this case a bifurcation appears at $\frac{\lambda}{\delta} > \frac{1}{\sqrt{N-1}}$. Finally, the third case regards the stability of the upper solution, the ``$-$'' solutions of Eq.~\ref{eq:ss_star_sol}, which is a new characteristic of our model. Interestingly, the eigenvalues of $\J$, in this case, are also both negative for any set of parameters with $\lambda^* > 0$. The calculation of the eigenvalues of $\J$ can be done analytically as well as the proof that the solutions are stable (negative eigenvalues). However, this is not shown since the expressions are too big and have no interesting information.

\begin{figure}[t]
\includegraphics[width=\linewidth]{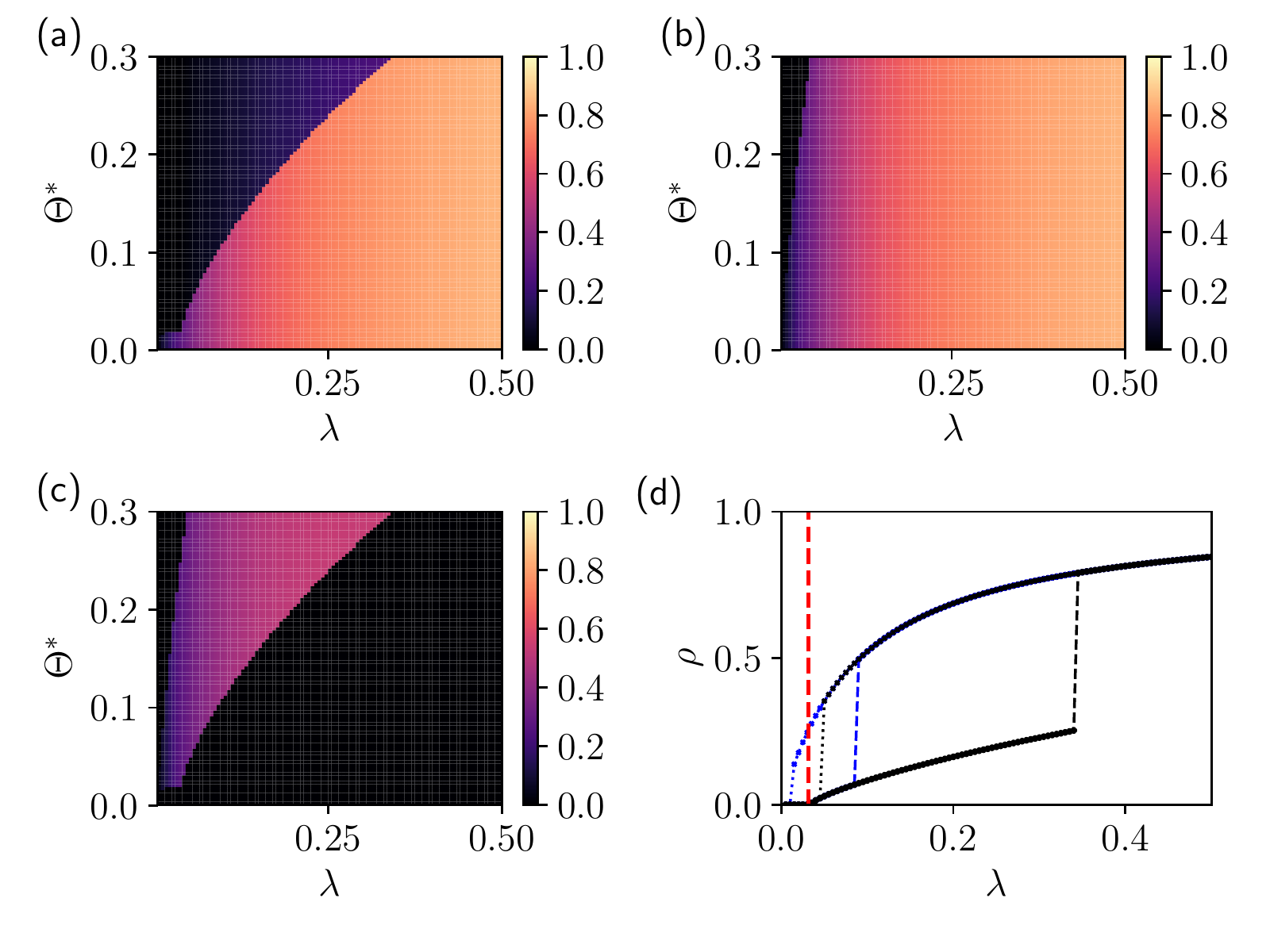}
\caption{Phase diagram for the star hypegraph with $N = 10^3$, $\delta = 1$ and $\lambda^*(|e_j|) = \log_2(|e_j|)$. In (a)-(c) the colormaps are obtained changing $\lambda$ and $\Theta^*$. In (a) the solution of Eq.~\ref{eq:star} from $y_c = y_l = 0.01$ as initial condition, in (b) the solution of Eq.~\ref{eq:star} from $y_c = y_l = 1.0$ and in (c) the latent heat (i.e. difference between (b) and (a)), emphasizing the hysteresis loop. In (d) the phase diagram for $\Theta^* = 0.1$ in blue and $\Theta^* = 0.3$ in black, where the dashed lines have $y_c = y_l = 0.01$ as initial conditions and dotted lines have $y_c = y_l = 1.0$. The red dashed line is the critical point prediction for a star graph using the QMF, $\lambda_c = \left(\sqrt{N-1}\right)^{-1}$.}
\label{fig:application_1}
\end{figure}

\subsection{Critical values}

The critical values are obtained as the points when $\Theta^*$ crosses $\rho^{\text{Lower}}$ and $\rho^{\text{Upper}}$ for the lower and upper solution respectively. Formally,
\begin{eqnarray}
 &\lambda_c^{\text{L}} =& \text{arg}_{\lambda} \left( \rho^{\text{Lower}}(\lambda, \delta, \lambda^*, N) = \Theta^* \right) \\
 &\lambda_c^{\text{U}} =& \text{arg}_{\lambda} \left( \rho^{\text{Upper}}(\lambda, \delta, \lambda^*, N) = \Theta^* \right),
\end{eqnarray}
where, we are assuming that $\delta$, $\lambda^*$ and $N$ are kept fixed and the critical point is calculated in terms of $\lambda$. Note that more general expressions might be calculated using the same principle. In the star case, this expression was not obtained analytically, but this can be calculated numerically.

\subsection{``Latent heat''}

As previously mentioned, the latent heat can be expressed as
\begin{equation}
 Q_l(\lambda_c^{X}) = \left( \rho^{\text{Upper}}(\lambda, \delta, \lambda^*, N) - \rho^{\text{Lower}}(\lambda, \delta, \lambda^*, N) \right)_{\lambda = \lambda_c^X}
\end{equation}
where $Q_l(\lambda_c^{X})$ can be $Q_l(\lambda_c^{\text{L}})$ or $Q_l(\lambda_c^{\text{U}})$. This equation can be analytically expressed, but its is too long. Conversely, one might analyze this quantity in the thermodynamic limit. Formally, it is expressed as
\begin{equation}
 \lim_{N \rightarrow \infty} Q_l(\lambda_c^{X}) = \left( \frac{\delta  \lambda^*  \lambda }{(\delta +\lambda ) (\delta +\lambda^*  \lambda +\lambda )} \right)_{\lambda = \lambda_c^X}.
\end{equation}

\subsection{Finite-size effects: analytical results vs ODE solutions}

In Fig.~\ref{fig:application_1} we show the phase diagram for the hyperstar with $N = 10^3$, $\delta = 1$ and $\lambda^*(|e_j|) = \log_2(|e_j|)$. This diagram was obtained solving the ODE system~\ref{eq:star}. It can be compared with Fig.~\ref{fig:analytical_pd}, which are analytical solutions of the same system. By comparing them, we can observe that there is a mismatch between both solutions as the threshold increases. More specifically, both the upper and lower solution perfectly fits our analytical results far from the discontinuities. The only observed problem is the position of the discontinuity in the lower solution.

\section{Upper solution: fluctuations and finite size effects} \label{sec:upper_sol_mc}

\begin{figure}[t]
\includegraphics[width=\linewidth]{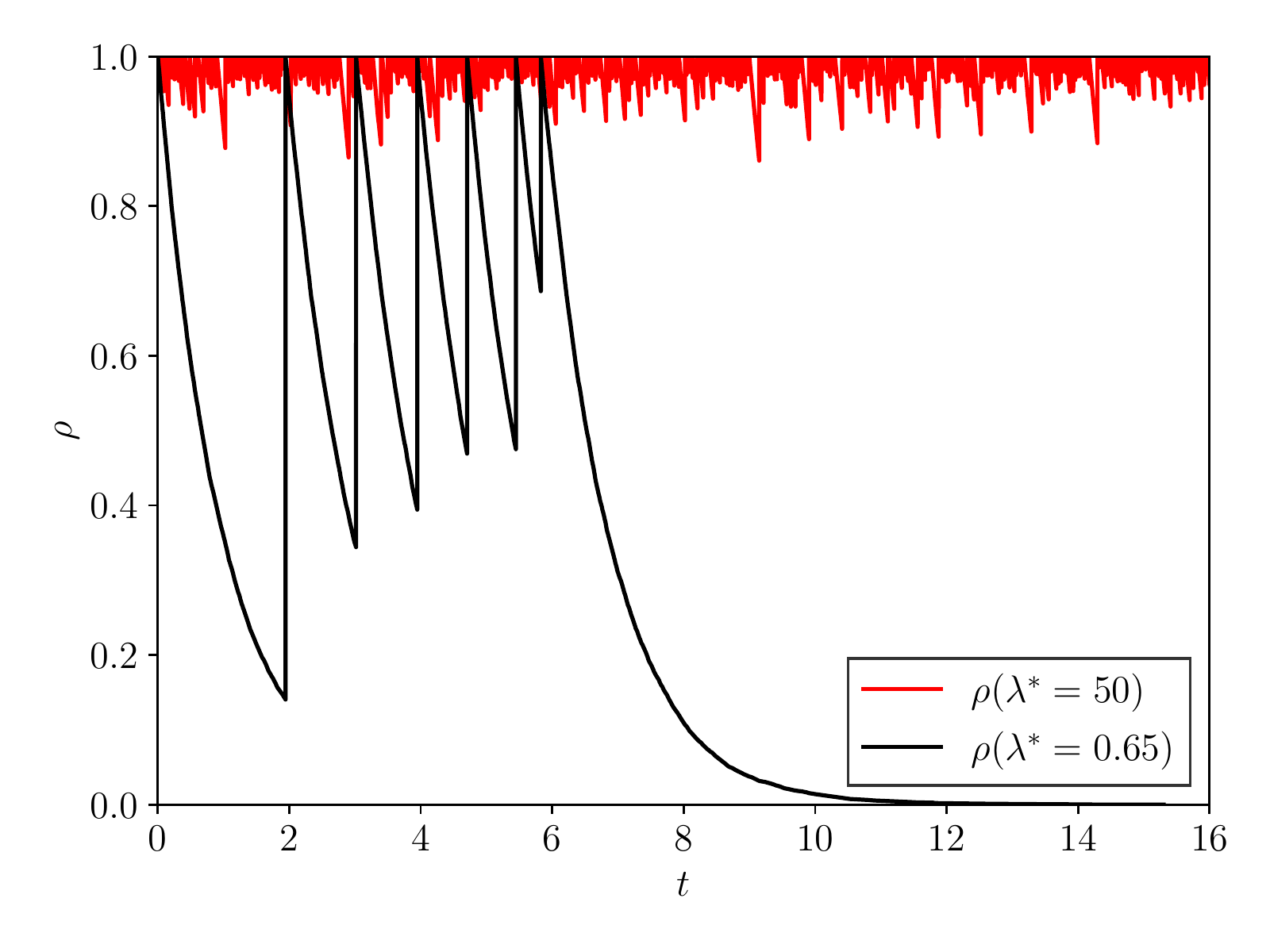}
\caption{Monte Carlo simulation considering a hypergraph composed by single hyperedge, $e_1 = \{1, 2, ..., N \}$, where $N = 10^4$, $\delta = \lambda = 1.0$ and two different values of $\lambda^* = 50$ and $\lambda^* = 0.65$. Each jump is a consequence of the hyperedge spreading. After the last jump we had an event where $t_s < t_r$, hence, allowing the dynamics to fall into the absorbing state.}
\label{fig:MC_t_Upper}
\end{figure}

In finite hypergraphs, due to fluctuations and as time goes to infinity, we expect that an upper solution might fall into a lower solution. Following the same reasoning, the lower solution might also fall into the absorbing state or even jump to the upper solution. Note that, the only absorbing state is $\rho = 0$; hence, in a finite hypergraph, in infinity time, the dynamics will always reach this state. We remark that the results we obtained in the previous sections neglect such fluctuations, accounting only for the average. To better understand this, we consider the simplified extreme case. Thus, defining a process where every node is disconnected by any lower-order interaction aside from the $|e_1| = N$. Thus, we have $N+1$ Poisson processes, $N$ deactivation processes, $N_i^\delta$, and one spreading $N_1^{\lambda \lambda^*}$. The latter process defines the characteristic spreading time, $t_s$, while the others define the time necessary to arrive at the critical mass, $t_r$. Since, by definition, all of them are Poisson processes,
\begin{equation}
 t_s = (\lambda \lambda^*)^{-1},
\end{equation}
while
\begin{equation}
 t_r = N (1 - \Theta^*) \delta^{-1},
\end{equation}
where $N (1 - \Theta^*)$ is the necessary number of nodes to maintain the process $N_1^{\lambda \lambda^*}$ active and $\delta^{-1}$ is the average time to deactivate a single process. From these results, in order to obtain a bound for the average time to reach $\rho \geq \Theta^*$ we should respect $t_s < t_r$. Hence,
\begin{equation} \label{eq:upper_cond}
 \frac{\lambda \lambda^*}{\delta} > \left(N (1 - \Theta^*) \right)^{-1}.
\end{equation}
This equation can be interpreted as a lower bound regarding the structure since we neglected all possible lower-order hyperedges. We also remark that it is a lower bound for the average and fluctuations can take the dynamics to the absorbing state, even respecting Eq.~\ref{eq:upper_cond}. 

Fig.~\ref{fig:MC_t_Upper} shows this behavior. We have two competing processes, the exponential decay is a consequence of the deactivation mechanism, while the abrupt activation of all nodes is given by the hyperedge spreading. When $\lambda^*$ is sufficiently large, the system takes longer to fall into the absorbing state. On the other hand, when it is relatively small, it oscillates, until reaching $\rho < \Theta^*$, when just the annihilation mechanisms exist and the system can not get to the meta-state anymore.  
We remark that this argument is valid only in finite systems. In the thermodynamic limit, the fluctuations are not able to move from one solution to another.

\section{Monte Carlo simulations} \label{sec:MC}

\subsection{Continuous-time simulations}

To statistically describe our model, we use the continuous-time Monte Carlo simulations. More specifically, we use the Gillespie algorithm~\cite{Gillespie1977} to implement the dynamics. The algorithm is described as follows. We create a vector with all possible Poisson processes. This vector contains the time in which the events are expected to happen. If the process is not active, we set it as $\infty$. If it is active, we sum the current time with a $\Delta t$ sampled from an exponential distribution with the proper parameter, i.e., given by the associated Poisson process. Thus, given an initial condition, the dynamics run on top of this vector. On each iteration, we find the element with the shortest time and execute its rule, which can be deactivation or spreading (pair-wise or in the hyperedge if the threshold is reached). This also implies that new processes might be created or deleted accordingly. Next, time is updated, and the same process is repeated until reaching the absorbing state or a $t_{max}$. This algorithm is a simple extension of the methods described in Section 10.3 of~\cite{Arruda2018}.

\subsection{Quasi-stationary method (QS)}

Aside from the dynamical evolution of our system, we also use the quasi-stationary method (QS) to avoid the absorbing state, obtaining a statistically reliable characterization of our process. This method was initially proposed in~\cite{Oliveira2005} and had been extensively used in the analysis of epidemic spreading~\cite{Ferreira2012, Mata2013, Arruda2018} The algorithm is defined as follows. We keep a list of $M$ previously visited active states. This list is continuously updated. If we are in an active state, with a probability $p_r \Delta t$ the current state replaces a random position of this list. If the absorbing state is reached, then a random element of the list replaces the absorbing state. In this way, the absorbing state is avoided. To obtain statistically meaningful results, we let the dynamics relax for $t_r$ and, after that, we sample the distribution of $\Prob{n_{active}}$ during a time $t_s$. Note that, on each iteration of the described algorithm, we are computing $\text{Freq}(n_{active}) \leftarrow \text{Freq}(n_{active}) + \Delta t$. Physically, we are computing the time our dynamics spent in the state $n_{active}$. Therefore, $\Prob{n_{active}} \propto \text{Freq}(n_{active})$. From that, we can describe our system using the order parameter and the susceptibility. Formally, they are respectively expressed as
\begin{eqnarray}
 \rho &=& \E{n_{active}}, \\
 \chi &=& \frac{\E{n_{active}^2} - \left(\E{n_{active}}\right)^2}{\E{n_{active}}},
\end{eqnarray}
where $n_{active}$ is the number of active nodes in the dynamics. We remark that, in a second-order phase transition, the susceptibility diverges in the thermodynamic limit. On the other hand, in a first-order phase transition, the susceptibility and the order parameter have discontinuities.

We remark that $t_r$ and $t_s$ vary according to the system size. On the other hand, the algorithm is stable to the choices of the size of the list $M$ and the probability $p_r$. In order to reduce the computational cost of this method, we also employed an adaptive version. In this version, we define a variable sampling time given as $t_r + c t_s^*$, where $t_s^*$ is a smaller time-window and $c$ is not set but defined by the convergence of $\chi$. In practice, we calculate $\chi$ before and after each $t_s^*$ time-window. If the absolute difference between the susceptibility is lower than $\epsilon$ (here set as $\epsilon = 0.001$) then the algorithm stops. Additionally, we also define a $c_{max}$ (here set as $c_{max} = 500$), which is the stop condition. Thus, with this adaptive version, we expect to reduce the computational cost but keeping statistically reliable measurements. 

The main computational challenge introduced by our model is the characterization of the bi-stable region. On the one hand, for the single-absorbing state, we have the QS method, which solves the difficulties introduced by the absorbing state. On the other hand, we have a region with two solutions. The simplest solution is to initialize the QS method with different initial conditions, $\rho(t=0) = 1.00$ and $\rho(t=0) = 0.01$, and obtain the desired statistics independently. In this manner, if the hypergraph is large enough, we expect that the time to jump from one solution to the other is sufficiently large. It is expected to be infinity in the thermodynamic limit, as already mentioned.

\subsection{Estimating the ``latent heat''} \label{sec:estimating}

In our experiments, we observed that near the discontinuities we might observe a peak in the susceptibility. This peak seems to be an artifact of the simulation in finite systems. For instance, considering the upper (lower) solution. Note that, at the discontinuity, due to too strong fluctuations can lead the system to the lower (upper) solution or even the absorbing state. This random event should be less and less likely to happen as we increase the system size. However, for practical reasons, the QS method has a maximum time stopping condition. Near the discontinuities, this jump between solutions is more likely than in other parts of the phase diagram. Therefore, the time spent in each solution seems to be also an artifact of the simulations. However, both the average, but especially the variance, will be affected by this effect. 

The QS method is able to properly characterize both the upper and lower solution of our dynamics. Therefore, we are also able to estimate the latent heat. To avoid the previously mentioned artifact, we should measure the latent heat as
\begin{equation} \label{eq:Ql_iter}
 Q_l^{QS}(\lambda_c^{\text{X}}) =
 \rho(\tilde{\lambda}_c^{\text{X}}+\epsilon) - \rho(\tilde{\lambda}_c^{\text{X}}-\epsilon'),
\end{equation}
where $X$ indicates lower or upper solution, $\tilde{\lambda}_c$ are the estimations of the critical point and the $\epsilon$'s should be refined in simulations. To do so, a first run with a more spaced values of $\lambda$, obtaining a first guess of $\tilde{\lambda}_c^{\text{X}}$. Next, it should be fine tuned using Eq.~\ref{eq:Ql_iter}. Finally, $Q_l^{QS}(\lambda_c^{\text{X}})$ should be estimated guaranteeing that $\rho(\tilde{\lambda}_c^{\text{X}}+\epsilon)$ is in the upper solution and $\rho(\tilde{\lambda}_c^{\text{X}}-\epsilon')$ is in the lower solution. Aside from that, the discontinuity point is estimated as
\begin{equation}
 \lambda_c^{\text{X}} = \frac{1}{2} \left( 2 \tilde{\lambda}_c^{\text{X}}+\epsilon - \epsilon' \right).
\end{equation}
Importantly, as we refine our simulations, we might also increase the sampling and relaxation times, $t_s$ $t_r$, in the QS method. This allow us to obtain a better statistics of our system. Next, for comparison, we can use this estimation of $\lambda_c^{\text{X}}$ as an input to our analytical expressions of latent heat.

\subsection{Hysteresis and upper solution particularities}

Following from the discussion in Section~\ref{sec:upper_sol_mc}, the upper solution might also be harder to characterize. For instance, the regular cases studied have a hyperedge of the size of the whole population. Thus, they are expected to present a similar behavior as depicted in Fig.~\ref{fig:MC_t_Upper}. Indeed, due to this ``sawtooth-like'' behavior, simulating the dynamics for a fixed amount of time and computing $\rho (t = t_{\max})$ for several runs will show the variance associated with our process in this structure. Note that, for more complex structures, the solutions cannot be reduced to this ``sawtooth-like'' behavior.

Complementary to the QS method, we also employ a simpler algorithm to study the phase diagram. We run the dynamics independently $N_{\text{runs}} = 50$, runs for $t_{\max} = 100$ and calculate only the final fraction of active nodes, $\rho(t_{\max})$. This algorithm has a much lower computational cost if compared with the QS since we limit the $t_{\max}$ to a much lower value. If compared with the QS, it is less likely that the upper solution will fall into the lower due to stochastic fluctuations. Finally, since we do not interfere in the temporal course of the process, this can be directly compared with the ODE's numerical solution, which is the main advantage of this method.

\subsection{Regular hypergraphs: Hyperblob and Hyperstar} \label{sec:QS_regular}

\begin{figure}[t]
\includegraphics[width=\linewidth]{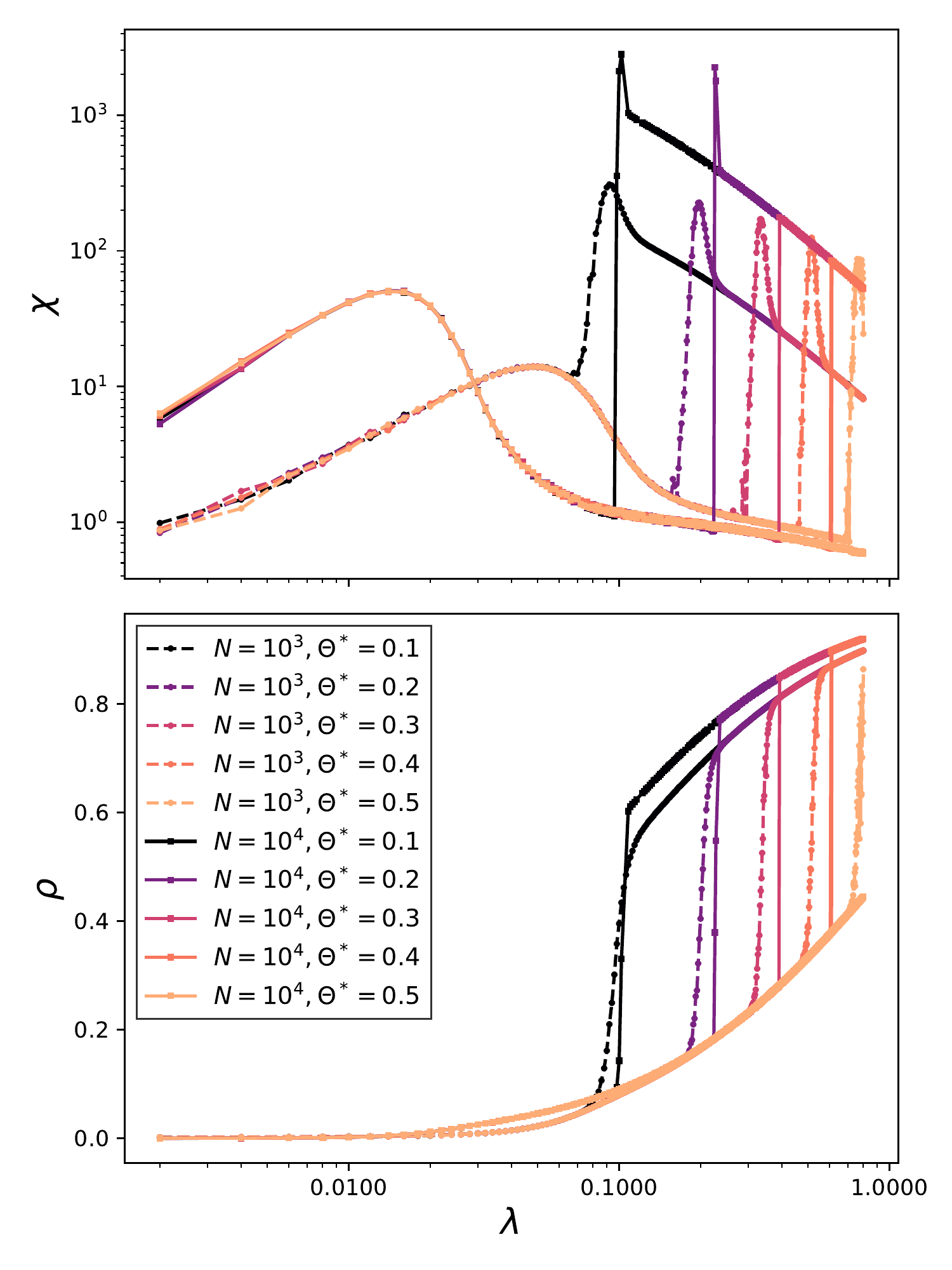}
\caption{Estimation of $\rho$ and $\chi$ using the QS method in a
Hyperstar with $\delta = 1.0$ and $\lambda^* = \log_2(|e_j|)$ for
different sizes and critical-mass thresholds. In the top panel we
present the the susceptibility, while on the bottom panel the order
parameter.}
\label{fig:HyperStar_QS}
\end{figure}

\begin{figure}[t]
\includegraphics[width=\linewidth]{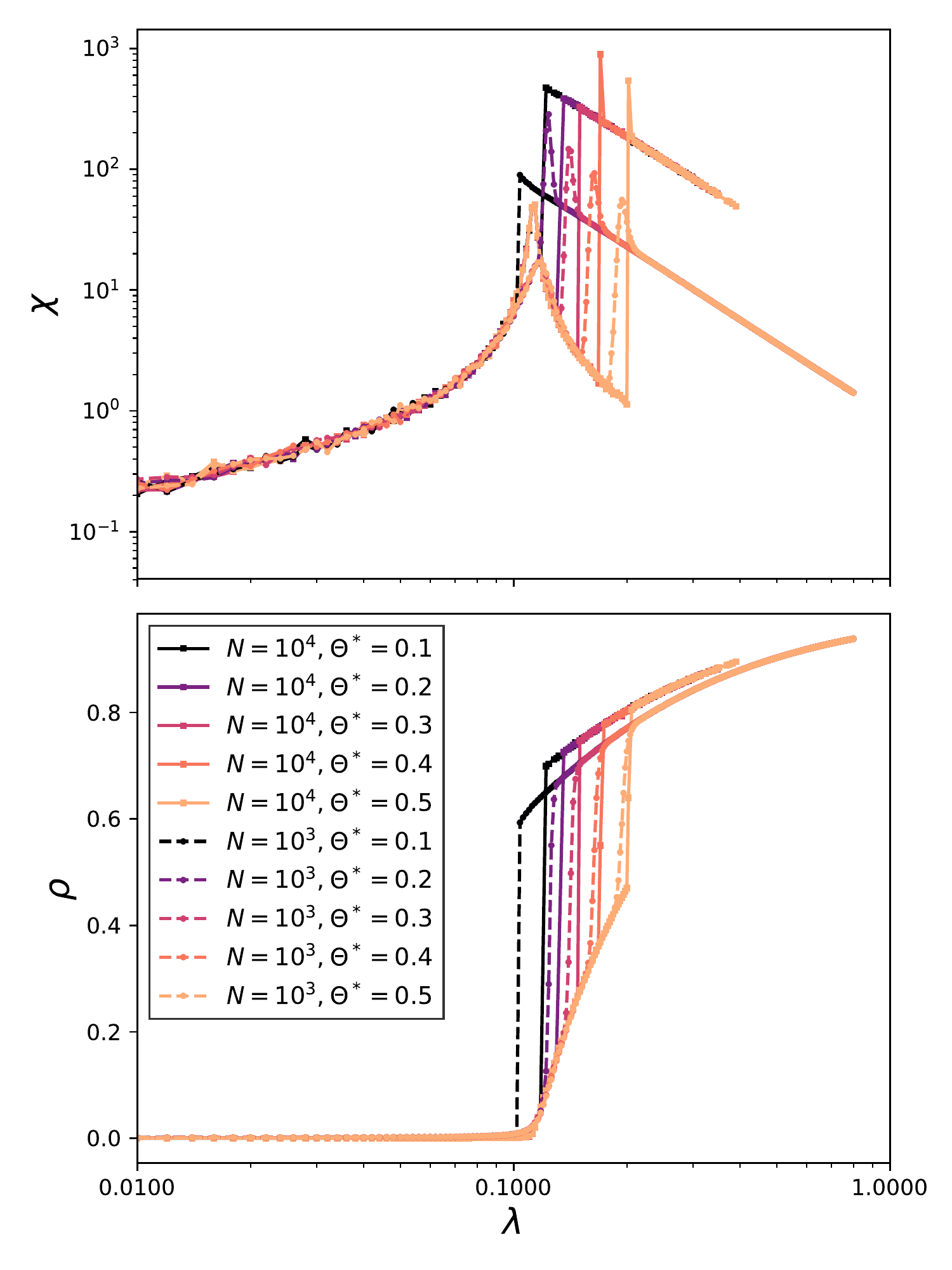}
\caption{Estimation of $\rho$ and $\chi$ using the QS method in our
regular homogeneous hypergraph with $\delta = 1.0$ and $\lambda^* =
\log_2(|e_j|)$ for different sizes and critical-mass thresholds. In
the top panel we present the the susceptibility, while on the bottom
panel the order parameter.}
\label{fig:Hyper_RRN}
\end{figure}

\begin{table}[t]
\caption{Discontinuity point and latent heat estimations for the
hyperstar and the homogeneous hypergraph with $N = 10^4$. The fixed
dynamical parameters are $\delta = 1.0$ and $\lambda^* =
\log_2(|e_j|)$. For different values of $\Theta^*$ we report the
discontinuity point, $\tilde{\lambda}_c^{\text{Lower}}$, the
calculated latent heat, $Q_l(\tilde{\lambda}_c^{\text{Lower}})$ (from
the expressions derived in Section~\ref{sec:Hyperstar}), the estimated
latent heat, $Q_l^{QS}(\tilde{\lambda}_c^{\text{Lower}})$ (see
Fig.~\ref{fig:HyperStar_QS}) and the absolute and relative errors,
which are calculated as $\epsilon_{abs} = |
Q_l(\tilde{\lambda}_c^{\text{Lower}}) -
Q_l^{QS}(\tilde{\lambda}_c^{\text{Lower}})|$ and $\epsilon_{rel} =
\frac{\epsilon_{abs}}{Q_l(\tilde{\lambda}_c^{\text{Lower}})}$}
\label{tab:latent_heat}
\begin{tabular}{|c|c|c|c|c|c|}
\hline
$\Theta^*$ & $\tilde{\lambda}_c^{\text{Lower}}$ &
$Q_l(\tilde{\lambda}_c^{\text{Lower}})$ &
$Q_l^{QS}(\tilde{\lambda}_c^{\text{Lower}})$ & $\epsilon_{abs}$ &
$\epsilon_{rel}$ \\ \hline
\hline
\multicolumn{6}{|c|}{Hyperstar} \\
\hline
$0.1$ & $0.103$ & $0.502$ & $0.508$ & $6.54 \times 10^{-3}$ & $1.30
\times 10^{-2}$ \\ \hline
$0.2$ & $0.230$ & $0.580$ & $0.587$ & $7.30 \times 10^{-3}$ & $1.26
\times 10^{-2}$ \\ \hline
$0.3$ & $0.391$ & $0.567$ & $0.568$ & $8.63 \times 10^{-4}$ & $1.52
\times 10^{-3}$ \\ \hline
$0.4$ & $0.611$ & $0.518$ & $0.518$ & $5.85 \times 10^{-4}$ & $1.13
\times 10^{-3}$ \\ \hline
\hline
\multicolumn{6}{|c|}{Homogeneous hypergraph} \\
\hline
$0.1$ & $0.120$ & $0.543$ & $0.640$ & $9.66 \times 10^{-2}$ & $1.78
\times 10^{-1}$\\ \hline
$0.2$ & $0.133$ & $0.485$ & $0.570$ & $8.49 \times 10^{-2}$ & $1.75
\times 10^{-1}$ \\ \hline
$0.3$ & $0.149$ & $0.428$ & $0.479$ & $5.14 \times 10^{-2}$ & $1.20
\times 10^{-1}$ \\ \hline
$0.4$ & $0.167$ & $0.370$ & $0.425$ & $5.50 \times 10^{-2}$ & $1.49
\times 10^{-1}$ \\ \hline
$0.5$ & $0.202$ & $0.307$ & $0.340$ & $3.31 \times 10^{-2}$ & $1.08
\times 10^{-1}$ \\ \hline
\end{tabular}
\end{table}

Both the homogeneous and the hyperstar are expected to present extreme fluctuations, as initially discussed in
Section~\ref{sec:upper_sol_mc}. In fact, our simulations in these
systems, considering finite hypergraphs, allowed us only to
characterize $\rho^{\bigstar}$. This suggests that the bi-stable region only exists in the thermodynamic limit.

Fig.~\ref{fig:HyperStar_QS} shows the estimation of $\rho$ and $\chi$ using the QS method in a hyperstar. We observed that the second-order phase transition is well estimated, as can be seen by the rounded peaks. Note that, as the system size increases, this transition also moves to the left, as expected. Besides, it also seems to diverge, as predicted by a second-order phase transition. The difference from the hyperstar and a simple star graph is the discontinuity in the order
parameter and susceptibility curves. Fig.~\ref{fig:HyperStar_QS} also
suggests that in this discontinuity, the limit from both sides are not
the same, as expected. At the critical point, we observe an apparent
divergence of the susceptibility. However, this might be an artifact
of the simulations since the time we sample each solution will determine it. These effects are given by a stochastic factor, but also by the simulation parameters. Furthermore, the jump is already apparent for $N = 10^4$. For $N = 10^3$, we already see signs of a discontinuous transition, but it is not as clear as for larger system sizes.

Complementary, in Fig~\ref{fig:Hyper_RRN}, we show the estimation of
$\rho$ and $\chi$ using the same method. Note that, for larger values of $\Theta^*$, the susceptibility curve is very similar to the one observed by a random regular graph, as can be seen in~\cite{Mata2013}. Note that, as we increase the system size, this quantity also seem to diverge, as expected for a second-order phase transition. Next, as predicted, we also observed the jumps from the lower solution to the upper solution. Interestingly, in hypergraphs with $N = 10^3$, fluctuations before reaching the critical mass were sufficiently large to move the system to the upper solution. This behavior is particularly clear for $\Theta^* = 0.1$, where the system jumps even before the second-order phase transition.

In Table~\ref{tab:latent_heat} we present the estimations for the
latent heat and the discontinuity point as well as their absolute and
relative errors. In this table, we considered the hyperstar and the
homogeneous hypergraph with $\lambda^* = \log_2(|e_j|)$ and a variety
of critical mass thresholds, $\Theta^*$. The estimations of the latent
heat followed the procedure exposed in Section~\ref{sec:estimating}.
We observe that, although we performed a reasonable strong
approximation, the estimated latent heat is remarkably good for the
hyperstar case. The quantified error is around $O(10^{-3})$ in
absolute terms and around $O(10^{-2})$ in relative terms. In the
homogeneous hypergraph case, the estimations were poorer but still
reasonably good. The quantified errors, in this case, were around
$O(10^{-2})$ in absolute terms and around $O(10^{-3})$ in relative
terms.

\subsection{Heterogeneous cases: Exponential and Power-law cardinality distribution} \label{sec:MC_Exp_PL}

\begin{figure*}[t]
\includegraphics[width=\linewidth]{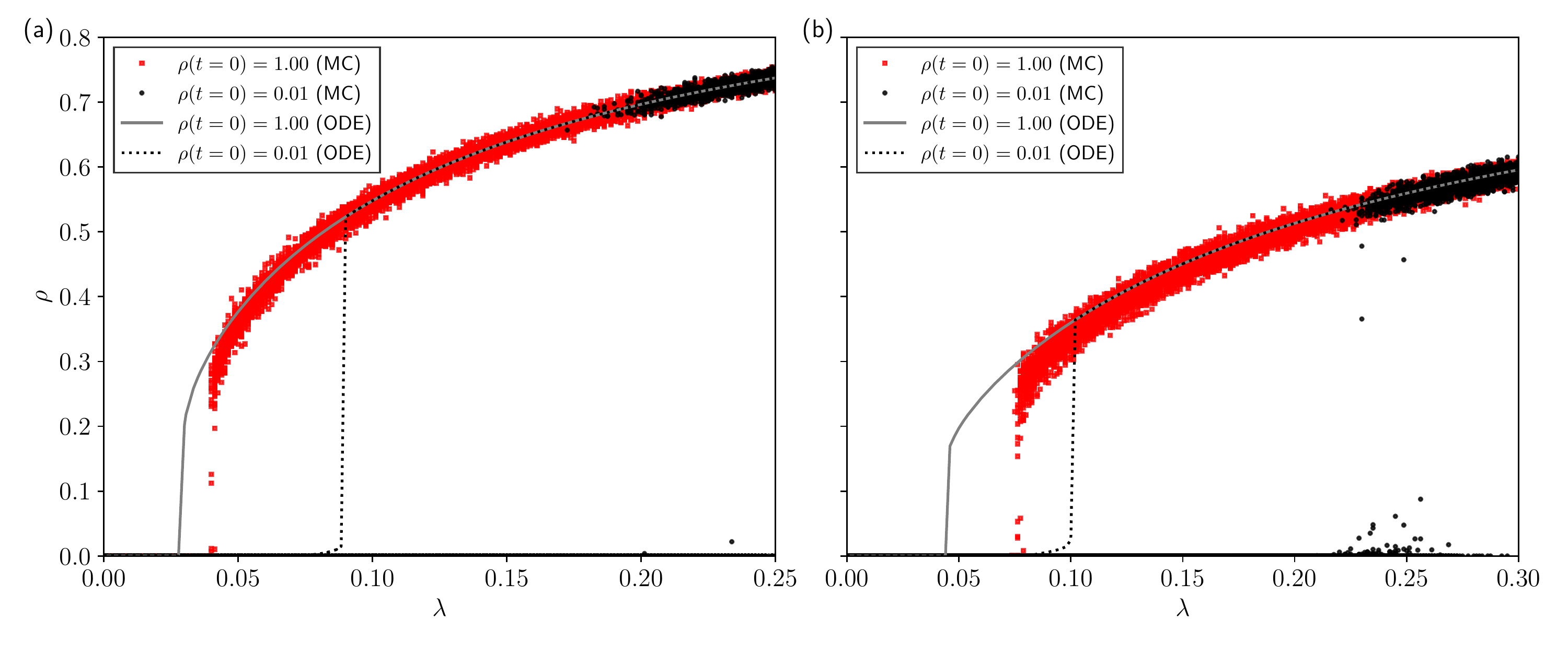}
\caption{Monte Carlo simulations for the exponential case in (a) and the power-law case in (b). Both hypergraphs have $N = 10^4$ nodes. The dynamical parameters are $\delta = 1.0$, $\lambda^* = \log_2(|e_j|)$, $\Theta^* = 0.2$. The initial conditions are color coded: $\rho(t=0) = 1.00$, in black, and $\rho(t=0) = 0.01$ in red.}
\label{fig:MC_PD}
\end{figure*}

\begin{figure*}[t]
\includegraphics[width=\linewidth]{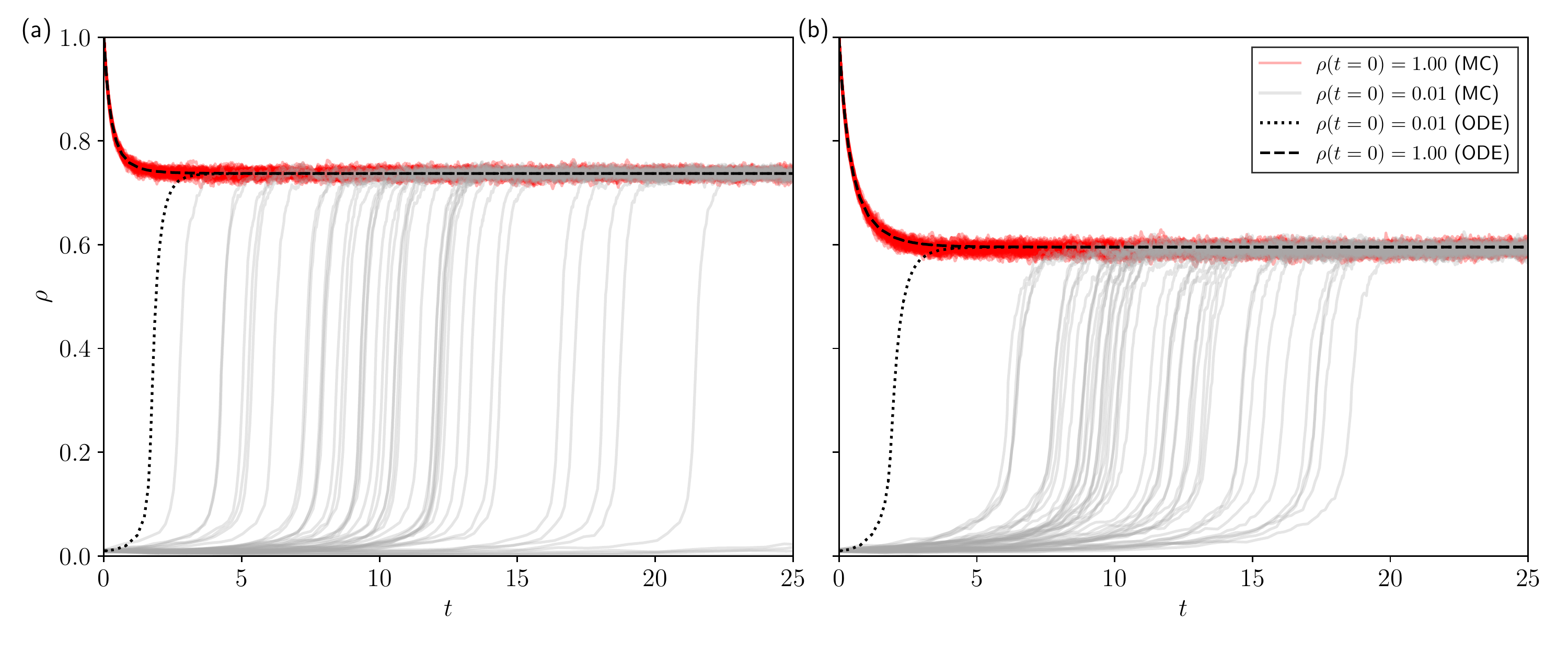}
\caption{50 runs of Monte Carlo simulations for the exponential case, with $\lambda = 0.25$ in (a) and the power-law case,  with $\lambda = 0.3$ in (b). Both hypergraphs have $N = 10^4$ nodes. The dynamical parameters are $\delta = 1.0$, $\lambda^* = \log_2(|e_j|)$, $\Theta^* = 0.2$. The initial conditions are color coded: $\rho(t=0) = 1.00$, in red, and $\rho(t=0) = 0.01$ in gray. The ODE's numerical solutions are also reported, for $\rho(t=0) = 1.00$, the dashed lines, and for $\rho(t=0) = 0.01$ the dotted lines.}
\label{fig:MC_ODE_time}
\end{figure*}

\begin{figure}[t]
\includegraphics[width=\linewidth]{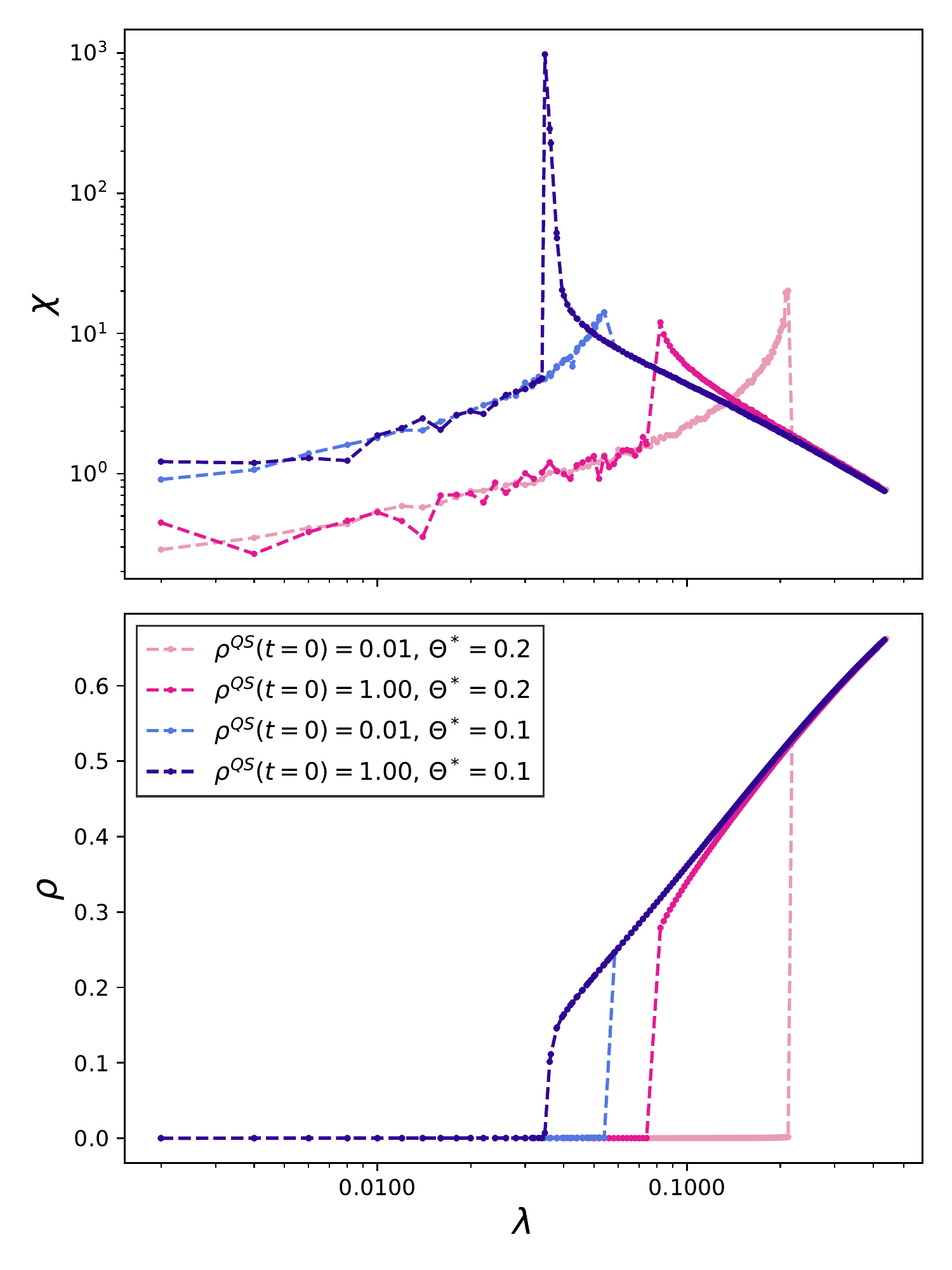}
\caption{Estimation of $\rho$ and $\chi$ using the QS method in a hypergraph with an power-law distribution of hyperedge cardinalities and $N = 10^4$. The dynamical parameter are: $\delta = 1.0$, $\lambda^* = \log_2(|e_j|)$ and $\Theta^* = 0.2$. In the top panel we present the the susceptibility, while on the bottom panel the order parameter. We considered two initial conditions for the QS method, $\rho^{QS}(t=0) = 0.01$ in black and $\rho^{QS}(t=0) = 1.00$ in red.}
\label{fig:HyperPL_QS}
\end{figure}

\begin{figure}[t]
\includegraphics[width=\linewidth]{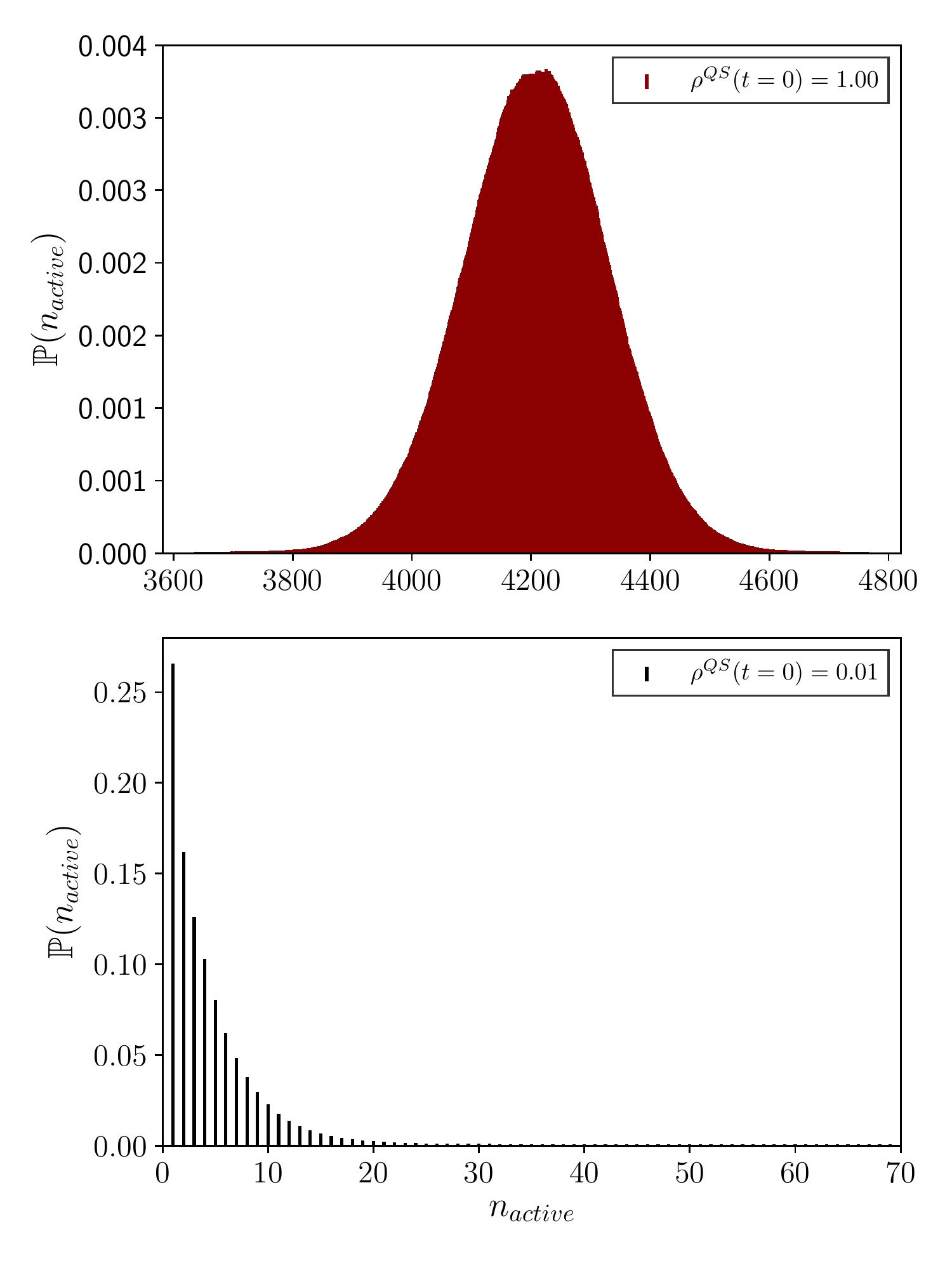}
\caption{Distribution of active node estimated using the QS method. The is same hypergraph as in Fig.~\ref{fig:HyperPL_QS}. he dynamical parameter are: $\delta = 1.0$, $\lambda = 0.138$ (the crossing between the two susceptibility curves, see Fig.~\ref{fig:HyperPL_QS}), $\lambda^* = \log_2(|e_j|)$ and $\Theta^* = 0.2$. In the top panel we present the distribution obtained using $\rho^{QS}(t=0) = 1.00$, while in the bottom pannel for $\rho^{QS}(t=0) = 0.01$.}
\label{fig:HyperPL_Drho_QS}
\end{figure}

Fig.~\ref{fig:MC_PD} shows the Monte Carlo simulations and ODE numerical solution for both exponential, in (a), and power-law cardinality distribution, in (b). The bi-stability region (hysteresis) is clear either from the simulation and from the numerical solutions. In both structures we observed that $\lambda_c^{\text{U}}$ is systematically better predicted than $\lambda_c^{\text{L}}$. It suggests that the odes well characterize $\lambda_c^{\text{U}}$. On the other hand, the existence of $\lambda_c^{\text{L}}$ is correctly predicted but poorly estimated, suggesting that further analysis should be carried out. Aside from that, we also observe that a more homogeneous structure, here the exponential case,showed a more accurate estimation. Despite that, the variance is also slightly smaller for the exponential structure than for the power-law.

The analysis of regular cases suggests that the lower solution is associated with the activation of lower cardinality hyperedges, while the upper solution is associated with the activation of higher cardinalities. The definition of our system also suggests this. Fig.~\ref{fig:MC_PD} suggests that the lower solution in both cases is the absorbing state or very close to it. It is only different from zero near the discontinuity. Thus, it is instructive to evaluate this structure separately. Although the highest probability of hyperedge cardinality is $\Prob{|e_j| = 2}$, the giant connected component considering only pairwise interactions is usually minimal. Note that in our model, hyperedges are created without no preference. In our specific case, for the exponential cardinality distribution, this connected component has six nodes, while for the power-law case, it has 29. Therefore, the higher-order hyperedges are responsible for most of the dynamics. 

Complementary, in Fig.~\ref{fig:MC_ODE_time}, we show the temporal behavior of our model. Similar conclusions also apply here. The upper initial condition has a better correspondence with the simulations. The lower initial condition presents a higher variance in terms of the time necessary to achieve the meta-state. We observe that the time it takes to get to the meta-state seems to depend on the initial micro-state as well as stochastic factors. Furthermore, note that in both experiments, the numerical solutions tend to predict this change earlier than it was observed in our simulations. 

Next, using the QS method, we can precisely determine the quantities of interest, such as the critical points and latent heat. These quantities could have been estimated using previous figures, but the estimation is expected to be less robust. In Fig.~\ref{fig:HyperPL_QS}, we present the estimations for the order parameter and the susceptibility using the QS method. Contrasting with the regular cases studied in the previous section, here we can observe the hysteresis. Furthermore, susceptibility reveals new phenomena. Aside from the discontinuity, also observed in the regular cases, here we can find two curves for the susceptibility. Both curves present a discontinuity at the same point as the order parameter. Their behavior can be understood evaluating the distribution of active nodes, shown in Fig.~\ref{fig:HyperPL_Drho_QS} for $\lambda = 0.138$, which is approximately the crossing point between the two susceptibility curves. The lower solution is dominated by the absorbing state and fluctuations around it. Note that, by construction $\Prob{n_{active} = 0} = 0$. On the other hand, in the upper solution, it is a bell-shaped distribution. It is instructive, to emphasize that a similar bell-shaped distribution is also expected above the threshold in an SIS epidemic spreading dynamics~\cite{Mieghem2009}.

For the sake of completeness, we recall that the analysis of the susceptibility for the exponential cardinality distribution hypergraph was shown in the main text.

\section{Implementation details}

The simulations were implemented in C/C++ using the standard libraries. The random numbers were extracted using the Gnu Scientific library~\cite{galassi2018scientific}. The ODE solutions were implemented using the Gnu Scientific library~\cite{galassi2018scientific}. More specifically, we used the explicit embedded Runge-Kutta-Fehlberg (4, 5) method, with an adaptive step-size control, where we keep the local error on each step within an absolute error of $\epsilon_{abs} = 10^{-4}$ and relative error of $\epsilon_{rel} = 10^{-3}$ with respect to the solution $y_i(t)$. We also remark that in Fig.~\ref{fig:application_1}, we observed convergence difficulties in some points. To solve this issue, the imposed absolute and relative errors were reduced, allowing the proper convergence. Furthermore, Gnu Parallel~\cite{Tange2011} was also used to run many instances of the same code, both for the simulations or the numerical solutions.


\end{document}